\documentclass[reprint,superscriptaddress,nofootinbib,amsmath,amssymb,aps]{revtex4-2}

\usepackage{graphicx}%
\usepackage{dcolumn}%
\usepackage{bm}
\usepackage{tabularx}
\usepackage{appendix}
\usepackage{graphicx}
\usepackage{subcaption}
\usepackage{colortbl}
\usepackage{color}
\usepackage{cancel}
\usepackage{comment}
\usepackage[countmax]{subfloat}
\usepackage{framed}
\usepackage{mdframed}
\usepackage{multirow}
\usepackage{slashed}
\usepackage{booktabs} 
\usepackage{hyperref}
\usepackage{caption}
\usepackage{ragged2e}
\DeclareCaptionJustification{justified}{\justifying}
\usepackage{physics}
\usepackage{comment}
\usepackage{mathrsfs}
\usepackage{ulem}

\newcolumntype{L}[1]{>{\hsize=#1\hsize\raggedright\arraybackslash}X}%
\newcolumntype{R}[1]{>{\hsize=#1\hsize\raggedleft\arraybackslash}X}%
\newcolumntype{C}[2]{>{\hsize=#1\hsize\columncolor{#2}\centering\arraybackslash}X}%

\begin{document}

\begin{flushright}
YGHP-23-03, RUP-23-15 
\end{flushright}
\vspace*{-1cm}
\title{String-wall composites winding around a torus knot vacuum in an axionlike model}

\author{Minoru Eto}
\email{meto@sci.kj.yamagata-u.ac.jp}
\affiliation{Department of Physics, Yamagata University, Kojirakawa, Yamagata 990-8560, Japan}

\affiliation{Research and Education Center for Natural Sciences, Keio University, 4-1-1 Hiyoshi, Yokohama, Kanagawa 223-8521, Japan}

\author{Takashi Hiramatsu}
\email{hiramatz@rikkyo.ac.jp}
\affiliation{Department of Physics, Rikkyo University, Toshima, Tokyo 171-8501, Japan}

\author{Izumi Saito}
\email{izumi@gfd-dennou.org}
\affiliation{Graduate School of Engineering, Nagoya Institute of Technology, Nagoya, 466-8555, Japan}

\author{Yuki Sakakihara}
\email{yuki.sakakihara@yz.yamagata-u.ac.jp}
\affiliation{Graduate School of Science and Engineering, Yamagata University, Yonezawa, Yamagata 992-8510, Japan}

\preprint{YGHP-23-03}

\begin{abstract}
We study a simple axionlike model with a charged scalar $\phi$ and a double-charged scalar $\zeta$ of global $U(1)$ symmetry. A particular feature of our model is that a vacuum manifold is a torus knot. We consider a hierarchical symmetry-breaking scenario where $\zeta$ first condenses, giving rise to cosmic $\zeta$-strings, and the subsequent condensation of $\phi$ leads to domain-wall formation spanning the $\zeta$-strings. We find that the formation of the walls undergoes two different regimes depending on the magnitude of an explicit breaking term of the relative $U(1)$ between $\zeta$ and $\phi$. One is the weakly interacting regime where the walls are accompanied by another cosmic $\phi$ strings. The other is the strongly interacting regime where no additional strings appear. In both regimes, neither a $\zeta$-string, a $\phi$-string nor a wall alone is topological, but the composite of an appropriate number of strings and walls %
as a whole is topologically stable,  characterized by the fundamental homotopy group of the torus knot.
We confirm the formation and the structure of the string-wall system by first-principle cosmological two-dimensional simulations. We find stable string-wall composites at equilibrium, where the repulsive force between $\zeta$-strings and the tension of walls balances, and a novel reconnection of the string-wall composites.

\end{abstract}
\date{\today}

\maketitle
\flushbottom

\section{Introduction}
\label{sec:introduction}

Topological defects are universal outcomes of phase transitions irrespective of the order of phase transition and 
one of the gravitational wave sources in the early Universe useful to probe high-energy physics~\cite{Auclair:2019wcv}.
Among a variety of topological defects, hybrid defects, composed of two different types of topological defects, e.g., strings bounded by monopoles and walls bounded by strings~\cite{Kibble:1982ae,Kibble:1982dd,Everett:1982nm,Preskill:1986kp,Martin:1996ea}, often appear in $SO(10)$ grand unification. It has been pointed out that the hybrid defects give unique gravitational wave signals~\cite{Chun:2021brv,Dunsky:2021tih}.
In the context of condensed-matter physics, walls bounded by strings are observed in superfluid ${}^3$He~\cite{Makinen:2018ltj}.

In this paper, we focus on and examine %
walls bounded by strings on field theoretical ground. 
We find such a configuration in high-energy physics as follows. 
One of the examples is $D$-parity, a $Z_2$ symmetry under the charge conjugation 
found in symmetry-breaking paths from grand unified theory (GUT) to Standard Model (SM). In \cite{Kibble:1982ae,Kibble:1982dd}, they pointed out that, %
in a symmetry-breaking path, 
\begin{align*}
    SO(10)\rightarrow SU(4)\times SU(2)_L\times SU(2)_R \times D \rightarrow G_{\rm SM} \ ,
\end{align*}
where $G_{SM}$ is the gauge group of Standard Model,
the first symmetry-breaking produces strings and then 
the second breaking produces domain walls (walls) spanning the strings.~\footnote{In these paths including $D$-parity breaking, there is a problem of the appearance of monopoles at the first symmetry-breaking and there is a necessity for some resolution of monopole dominance at the late time. Inflation is a resolution for it and recent works on cosmic strings suggest the replenishment of strings after inflation~\cite{Cui:2019kkd}.}
Another interesting example is the $U(1)_{B-L}$ extension of Standard Model. 
If gauged, $U(1)_{B-L}$ symmetry inevitably introduces right-handed neutrinos,
which are keys for producing neutrino masses and generating baryon-antibaryon asymmetry through leptogenesis~\cite{Buchmuller:2005eh}. 
The Higgs field giving masses to right-handed neutrinos is usually assigned $U(1)_{B-L}$ charge $2$, leading to $R$-parity as the remnant of $U(1)_{B-L}$. 
It was pointed out in \cite{Buchmuller:2007ui} that if we have, instead, a Higgs field charged $1$ under $U(1)_{B-L}$, $R$-parity breaks at the same time when the right-handed neutrinos obtain their masses and the gravitino, unstable but long-lived enough,  becomes a dark matter candidate. 
In this model, the breaking scale needs to be high enough to accommodate leptogenesis.
If we introduce both of the Higgs fields, physics related to right-handed neutrinos decouples from $R$-parity.
This scenario seems attractive since $R$-parity breaking-energy scale can be lowered, leaving us a larger parameter space for the leptogenesis.
This leads to a symmetry-breaking path down to Standard Model,
\begin{align*}
    G_{\rm SM}\times  U(1)_{B-L}
    \rightarrow G_{\rm SM} \times Z_2
    \rightarrow G_{\rm SM} \ ,
\end{align*}
ensuring the appearance of walls bounded by strings. 
We can understand these two examples in a unified way %
since, 1) in the first example, $D$-parity is the charge-conjugation symmetry which changes all the charges to the minus of the original ones and, in the second example, the remnant symmetry at the condensation of the double-charged field is the charge conjugation symmetry of the single-charged field and 2) the breaking of the charge-conjugation symmetry results in the formation of the walls attached to the strings in each example.

Testing these scenarios is difficult because very high energy is required to probe them. 
We would like to probe these scenarios through cosmological implications, e.g., the gravitational wave signals emitted from the string-wall systems
after the formation in the early Universe.
In this paper, as a first step, we explore the configuration and the dynamics of walls bounded by strings in the cosmological context in a simplest model with global symmetry,
\begin{align*}
    U(1) \rightarrow Z_2 \rightarrow I \ ,
\end{align*}
where $I$ is a trivial group. The first symmetry breaking leads to the string formation and the second breaking causes the formation of walls with the boundary of the strings previously formed.  
We realize this pattern by incorporating two $U(1)$ Higgs fields with different charges, $2$ and $1$~\cite{Krauss:1988zc,Ibanez:1991hv,Martin:1992mq}, similarly to the $U(1)_{B-L}$ models. We describe the impact of the gauge field on the configuration and the dynamics of the composite defects in another paper. We emphasize that our target in the present paper is qualitatively different from axionic domain walls examined before in Refs.~\cite{Vachaspati:1984dz,Ryden:1989vj,Battye:1994au,Khlopov:1999tm,Hiramatsu:2010yn,Hiramatsu:2012sc,Fleury:2015aca,Buschmann:2019icd,Buschmann:2021sdq}
in that the symmetry is 
not explicitly broken but spontaneously broken at the latter phase transition. 
This system has two complex scalar fields rather than one complex scalar field in the axionic case and has richer phenomena. 
It should be noted that our model is thought to be safe from the domain-wall problems even for $N > 1$ because of the appearance of the ends of the wall as we see in this paper.

In Sec.~II, we introduce our model, which includes two interacting fields with different charges under a global $U(1)$. %
In Sec.~III, we examine the vacuum structure at high energies and at low energies and explain the vacuum manifold is a torus knot in low energies. 
In Sec.~IV, we describe that there are two distinct regimes depending on the model parameters, where the potential structure is qualitatively different. Then, we explain nontopological defects found in our model and topological string-walls as composites of nontopological defects. We give examples of numerical solutions for string-walls in the flat background.
In Sec.~V, we perform cosmological simulations of our model and follow the time evolution of the string-wall systems for the two regimes. We also confirm the structure of the string-walls meets the prediction of the theoretical analysis in Sec.~IV.
Sections VI and VII are devoted to summary and discussion, respectively.
In Appendix~A, we show the mass formula for the four independent states around the vacuum.
In Appendix~B, we derive solutions for the walls in two regimes in our model. In Appendix~C, we explain how we generate the initial condition for the numerical simulations. In Appendix~D, we show enlarged figures of the reconnection happening in numerical simulation.

The metric is mostly positive, $(-,+,+,+)$ throughout the present paper.

\section{The Model}
\label{sec:the model}

We introduce two complex scalar fields, $\zeta$ and $\phi$,
in the Minkowski spacetime, $g_{\mu\nu}={\rm diag}(-1,1,1,1)$,
and define the action as
\begin{align}
    S&=\int d^4 x\; \sqrt{-g}\, \mathcal{L}_{\rm scalar}  \ ,
    \label{eq:action}\\
    \mathcal{L}_{\rm scalar}&=-\frac{1}{2} \partial_\mu \zeta^\ast \partial^{\mu} \zeta - \frac{1}{2} \partial_\mu \phi^{\ast}\partial^{\mu}\phi - V(\zeta, \phi)\ ,
    \label{eq:lagrangian}
\end{align}
with the potential 
\begin{align}
    V(\zeta, \phi)
    =& 
    \frac{\lambda_1}{4}(|\zeta|^2-v_1^2)^2+\frac{\lambda_2}{4}(|\phi|^2-v_2^2)^2 \nonumber \\
    &\; +\frac{m}{2} (\zeta^\ast \phi^2+  \phi^{\ast 2} \zeta )\ ,
    \label{eq:V}
\end{align}
where $m$ is a positive real parameter, determining the strength of the interaction between the two fields.~\footnote{Though $m$ can be a complex number, we assume $m$ is real and $m > 0$ without loss of generality since the phase of $m$ can be absorbed into the phase of the $\zeta$ field. 
}
We assume a hierarchical order 
\begin{eqnarray}
v_1 \gg \max\{v_2,m\} \ ,
\label{eq:hierarchy}
\end{eqnarray}
throughout this paper. 
We consider that the typical values are $v_1=10^{16}$ GeV, $v_2=10^2$ GeV and $\lambda_1=\lambda_2=1$, suggested by the NANOGrav recent result~\cite{NANOGRAV:2023gor,NANOGrav:2023ctt,NANOGrav:2023hvm,NANOGrav:2023hfp,NANOGrav:2023icp,Dunsky:2021tih,Maji:2023fba,Lazarides:2023ksx} and by GUT motivation, though the numerical simulations are performed for less hierarchical values for $v_1$ and $v_2$ because of the limitation of computational resources.

The action given by~\eqref{eq:action}-\eqref{eq:V} has been examined in different contexts,~\cite{Kim:2004rp,Choi:2014rja,Choi:2015fiu,Kaplan:2015fuy,Higaki:2016jjh,Long:2018nsl,Agrawal:2018mkd,Takahashi:2020tqv}. The difference between our work and these papers is whether there exists a hierarchy between the two condensation scales of two fields. 
If there is no hierarchy as in these papers, $v_1= v_2$, both of the strings of $\zeta$ and $\phi$ are formed at the same time, and the walls connect them by the explicit $U(1)$ breaking term in late time. In contrast, if $v_1 \gg v_2$ as in our study, only the strings of $\zeta$ are formed first, and the nontopological strings of $\phi$ attached to walls subsequently connect the strings of $\zeta$. 

In the following sections, we discuss two regimes depending on $m$, where the string-wall configuration is qualitatively different. The case with large $m$ has not been discussed in the previous works and is motivated by GUTs and (gauged) $U(1)_{B-L}$ models.
We investigate the torus knot vacuum structure and the string-wall configurations in the torus coordinate on the field space. We also solve the field equations and show that stable string-wall solutions exist.

\section{Vacuum structure}

\subsection{Symmetry}
The global symmetry in the absence of the third term of \eqref{eq:V} is $U(1)_\zeta \times U(1)_\phi$, which
is explicitly broken to a mixed $U(1)_{\rm mix}$ symmetry~\cite{Hiramatsu:2019tua,Hiramatsu:2020zlp} by the third term as
\begin{align}
\begin{split}
    \zeta & \rightarrow e^{i 2\theta } \zeta \ , \\
    \phi & \rightarrow e^{i \theta} \phi \ , 
\end{split}
    \label{eq:U(1)}
\end{align}
where $\theta$ is a real parameter of the $U(1)_{\rm mix}$ transformation.
In addition, there is a discrete symmetry
\begin{eqnarray}
    \mathscr{C}:~ (\zeta,\phi) \to (\zeta^*,\phi^*)\,.
\end{eqnarray}

First, $\zeta$ condensates, and then, the expectation values of $\zeta$ and $\phi$ become
\begin{eqnarray}
    \langle | \zeta | \rangle = v_1\,,\quad
    \left<\phi\right>=0\,.
\end{eqnarray}
The vacuum manifold 
at the higher energy scale is homeomorphic to $S^1$
whose fundamental homotopy group is
\begin{align}
    \pi_1(S^1)= \mathbb{Z} \ .
\end{align}
Note that it is parametrized by $\zeta = v_1 e^{2i\theta}$ with the $U(1)_{\rm mix}$ transformation parameter $\theta \in [0,2\pi]$. Hence, half-way of the $U(1)_{\rm mix}$ group manifold covers the full way of the vacuum manifold $S^1$. Note also that $U(1)_{\rm mix}$ is spontaneously broken to
$H_1=Z_2$ with $\theta = \{0,\pi\}$, while $\mathscr{C}$ remains unbroken.
Hence, the fundamental homotopy group can also be understood as 
\begin{align}
    \pi_1(G/H_1)=\pi_1(U(1)_{\rm mix}/Z_2)=\frac{1}{2} \mathbb{Z} \ ,
\end{align}
where $1/2$ means that the minimal unit is $\pi$ instead of $2\pi$ inside the $U(1)_{\rm mix}$ group manifold. %

When the second condensation occurs $\left<|\phi|\right> = v_2$ at lower energy, $H_1 = Z_2$ is spontaneously broken, so that we obtain a sequence of the symmetry breaking,
\begin{align}
    G=U(1) \xrightarrow{v_1} H_1=Z_2 \xrightarrow{v_2} I\,.
\end{align}
The second spontaneous symmetry breaking (SSB) provides a discrete structure in the vacuum manifold at low energies. However, the SSB as a whole is just $U(1) \to I$, and so, it is an issue how the vacuum manifold is structured. We will see that the vacuum manifold is a torus knot in the following subsections. On the other hand, the discrete symmetry $\mathscr{C}$ remains unbroken in the vacuum at any stage.

Since the $U(1)_{\rm mix}$ of our model is expressed as \eqref{eq:U(1)},
on any segment %
of a contour staying in a vacuum on the physical space, %
the following relation holds,
\begin{align}
 2 d\beta =  d \alpha \ ,
\label{eq:vac_rel}
\end{align}
where we define
\begin{align}
    \alpha := \arg (\zeta) \ , \qquad
    \beta := \arg (\phi) \ . %
\end{align}
To evaluate a winding number, we choose 
 $\alpha-\beta$, and the winding number is defined as
\begin{align}
 w:=\frac{1}{2\pi}\oint_{C_{\rm vac}} (d\alpha-d\beta) \ ,
\label{eq:winding}
\end{align} 
on a contour~$C_{\rm vac}$ staying in a vacuum on the physical space.
It can measure the winding since it behaves as
\begin{align}
 \alpha - \beta \quad \rightarrow  \quad \alpha - \beta + \theta
\end{align}
under \eqref{eq:U(1)}.~\footnote{There are also other combinations of $\alpha$ and $\beta$ to extract $\theta$.}
The field configurations with a vacuum boundary need to satisfy 
\begin{align}
 w = n \ , \quad (n \in \mathbb{Z}) \ .
\label{eq:vac_rel_int}
\end{align}

\subsection{Axionlike vacuum structure for $\alpha$-constant surface}
In order to figure out the structure of the low-energy potential, let us rewrite the scalar potential, Eq.~\eqref{eq:V},
\begin{eqnarray}
    V \simeq \frac{\lambda_2}{4}\left(|\tilde\phi|^2 - v_2^2\right)^2
    + \frac{m v_1}{2}\left(\tilde\phi^2 + \tilde\phi^{*2}    
    \right)\,,
    \label{eq:V_mod}
\end{eqnarray}
where we fixed $|\zeta| = v_1$ and absorbed the phase $\alpha = \arg \zeta$ by shifting the phase of $\phi$ as
\begin{eqnarray}
\tilde \phi = \phi\,e^{-i\alpha/2}\,.
\end{eqnarray}
This is formally the same as a scalar potential of the standard QCD axion model with $N=2$ domain walls.
The potential has two discrete symmetries
\begin{eqnarray}
    Z_2:~\tilde \phi \to -\tilde\phi\,,\qquad 
    \mathscr{C}':~\tilde \phi \to - \tilde \phi^*\,,
    \label{eq:discrete_sym}
\end{eqnarray}
where $\mathscr{C}'$ is given by a combination of $\mathscr{C}$ ($\tilde\phi\to\tilde\phi^*$) and $Z_2$ ($\tilde\phi \to - \tilde\phi$).
There are two vacua
\begin{eqnarray}
    \tilde \phi = \pm i \tilde v_2\ ,\quad 
    \tilde v_2 := v_2\sqrt{1+\frac{m}{m_\ast}}\ , \quad
    m_\ast:=\frac{\lambda_2 v_2^2}{2v_1} \,.
    \label{eq:two_vacua}
\end{eqnarray}
Thus, $Z_2$ is spontaneously broken, whereas $\mathscr{C}'$ remains unbroken. This is the reason why there exist two vacua.
Note that  $v_1 \gg \tilde v_2$ since we assume $v_1 \gg \max(v_2,m)$.

\subsection{Torus-knot vacuum structure}
 As we will explain below, these vacua are not discrete but are continuously connected. 
To illustrate the vacuum manifold, let us parametrize the fields $\zeta$ and $\phi$ as
\begin{eqnarray}
    \zeta = v_1 e^{i\alpha}\,,\qquad\phi = |\phi|e^{i\beta}\,,
\end{eqnarray}
with $\alpha,\beta \in [0,2\pi]$.
We fixed $|\zeta| = v_1$ and are left with three variables
$|\phi|$, $\alpha$, and $\beta$.
Then, the potential, Eq.~\eqref{eq:V}, becomes
\begin{eqnarray}
    V \simeq \frac{\lambda_2}{4}\left(|\phi|^2 - v_2^2\right)^2
    + m v_1|\phi|^2\cos\left(2\beta - \alpha\right)\,.
\end{eqnarray}

The potential minima are given by
\begin{align}
    |\phi| = \tilde v_2\,,
    \qquad
    2\beta - \alpha = \pm \pi\,.
    \label{eq:vac_cond}
\end{align} 
We found that the vacuum consists of the $Z_2$ discrete points, Eq.~\eqref{eq:two_vacua}, whereas it seems to be $S^1$ due to the condition 
$|\phi|=\tilde v_2$. 
Indeed, neither of them is the true vacuum structure.
\begin{figure}
    \centering
    \includegraphics[width=.85 \linewidth]{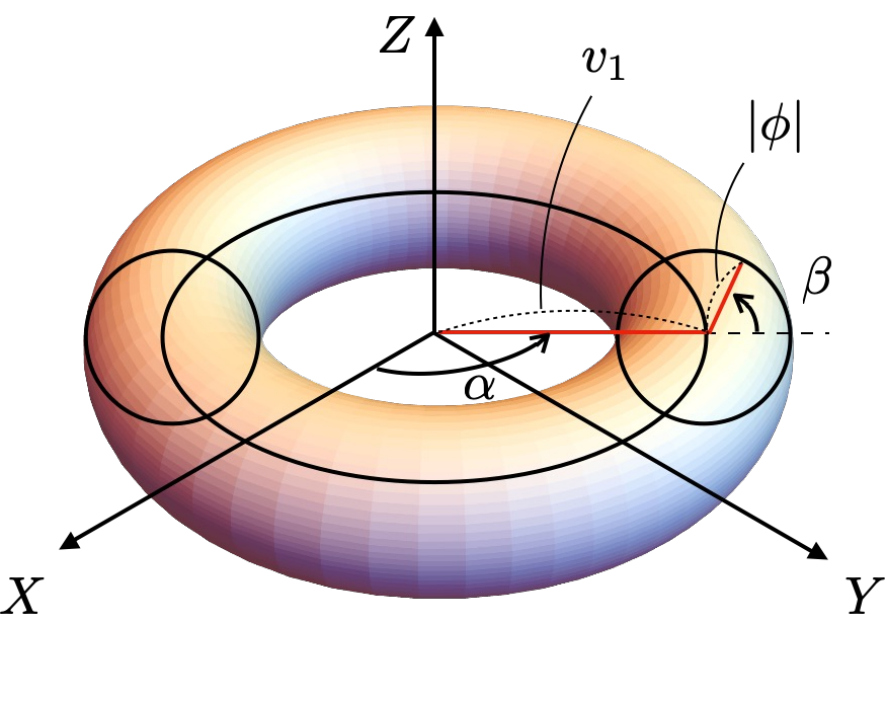}
    \caption{Field space mapped into the torus coordinate
    $(X,Y,Z)$ defined in Eq.~\eqref{eq:XYZ}
    The major and minor radii are $|\zeta|=v_1$ and $|\phi|$, respectively. The toroidal and poloidal angles are $\alpha$ and $\beta$, respectively.}
    \label{fig:XYZ}
\end{figure}
To uncover the vacuum structure, 
let us introduce new coordinate $(X,Y,Z)$ by
\begin{eqnarray}
    \left(
    \begin{array}{c}
    X\\
    Y\\
    Z
    \end{array}
    \right)
    = v_1\left(
    \begin{array}{c}
    \cos \alpha\\
    \sin\alpha\\
    0
    \end{array}
    \right)
    + |\phi|
    \left(
    \begin{array}{c}
    \cos\alpha \cos\beta \\
    \sin\alpha \cos\beta \\
    \sin\beta
    \end{array}
    \right).
    \label{eq:XYZ}
\end{eqnarray}
For a fixed $|\phi|$, as shown in Fig.~\ref{fig:XYZ}, $(X,Y,Z)$ represents the torus around the $Z$-axis with radii $v_1$ and $|\phi|$:
\begin{eqnarray}
    \left(\sqrt{X^2+Y^2}-v_1\right)^2+Z^2 = |\phi|^2\,.
\end{eqnarray}
Inversely, we can write down
\begin{align}
    \zeta &= v_1 \frac{X+iY}{\sqrt{X^2+Y^2}}\ , \\
    \phi &= \sqrt{X^2+Y^2}-v_1 + i Z\ .
\end{align}

The condition \eqref{eq:vac_cond} tells that the vacuum manifold is embedded in the torus with radii $v_1$ and $\tilde v_2$.~\footnote{When $m=0$, the vacuum manifold is the torus itself, which has two independent $S^1$s.}
Note that, since $v_1 \gg \tilde v_2$, the torus does not intersect itself.
The vacuum corresponds to the points on the torus satisfying
$2\beta - \alpha = \pm \pi$, which is represented as
two open curves parametrized by
$(\alpha,\beta) = (t,t/2\pm\pi)$ with $t \in [0,2\pi]$
\begin{eqnarray}
    \vec X_\pm (t)
    = v_1\left(
    \begin{array}{c}
    \cos t\\
    \sin t\\
    0
    \end{array}
    \right)
    + \tilde v_2
    \left(
    \begin{array}{c}
    \cos t \cos\left(\frac{t}{2}\pm\frac{\pi}{2}\right) \\
    \sin t \cos\left(\frac{t}{2}\pm\frac{\pi}{2}\right) \\
    \sin\left(\frac{t}{2}\pm\frac{\pi}{2}\right)
    \end{array}
    \right).
\end{eqnarray}
Note $\vec X_\pm(2\pi) \neq \vec X_\pm(0)$ but $\vec X_\pm(2\pi) = \vec X_\mp(0)$. Hence, we can smoothly connect the head of one curve and the tail of the other curve to form one closed curve. 
As a result, we can represent the vacuum manifold as
\begin{eqnarray}
    \vec X(t)%
    = v_1\left(
    \begin{array}{c}
    \cos t\\
    \sin t\\
    0
    \end{array}
    \right)
    + \tilde v_2
    \left(
    \begin{array}{c}
    \cos t \cos\left(\frac{t}{2}+\frac{\pi}{2}\right) \\
    \sin t \cos\left(\frac{t}{2}+\frac{\pi}{2}\right) \\
    \sin\left(\frac{t}{2}+\frac{\pi}{2}\right)
    \end{array}
    \right),
\end{eqnarray}
for $t=[0,4\pi]$.
When we vary $t$ from 0 to $2\pi$, the $\alpha$-cycle (the toroidal $S^1$ with the 
major radius $v_1$) winds once but the $\beta$-cycle (the poloidal $S^1$ with the minor  radius $\tilde v_2$) winds only one-half. The remaining $2\pi$ rotation from $t = 2\pi$ to $4\pi$ closes the curve and returns to the initial point. In total, the $\alpha$-cycle winds twice, and the $\beta$-cycle winds once. 
Hence, the vacuum manifold is the torus knot, which we refer to as ${\rm TN}(2,1)$, winding around a torus $\mathcal{T}$ defined by radii, $|\zeta|=v_1$ and $|\phi|=\tilde{v}_2$, 
\begin{eqnarray}
    \left(
    \begin{array}{c}
    X\\
    Y\\
    Z
    \end{array}
    \right)
    = v_1\left(
    \begin{array}{c}
    \cos \alpha\\
    \sin\alpha\\
    0
    \end{array}
    \right)
    + \tilde{v}_2
    \left(
    \begin{array}{c}
    \cos\alpha \cos\beta \\
    \sin\alpha \cos\beta \\
    \sin\beta
    \end{array}
    \right).
    \label{eq:torus_def}
\end{eqnarray}

\begin{figure}
    \centering
    \includegraphics[width=.95 \linewidth]{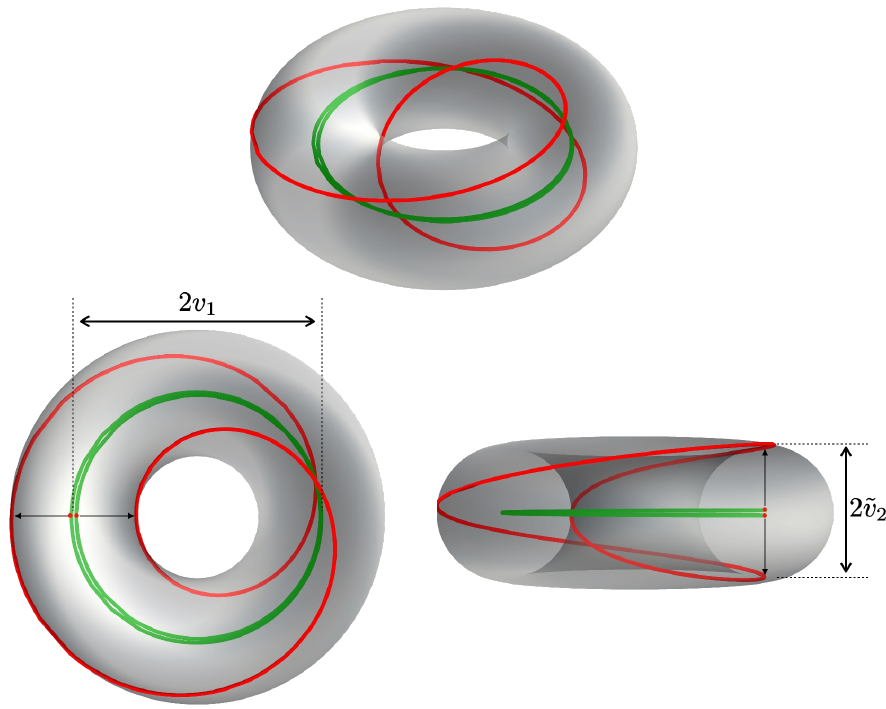}
    \caption{The red curve corresponds to the vacuum manifold TN(2,1) which lies on the auxiliary torus with the major and minor radii $v_1$ and $\tilde v_2$ in the $XYZ$-space. The green curve shows the vacuum manifold (doubly degenerate $S^1$ curves) before the second SSB takes place.}
    \label{fig:vacuum_manifold}
\end{figure}
In Fig.~\ref{fig:vacuum_manifold}, we show 
the vacuum manifolds at high energies ($\sim v_1 $) and low  energies ($\sim \tilde v_2$). The green curve corresponds to the vacuum manifold at high energies: the $S^1$ with the radius $v_1$ (the torus at the limit of 
$\tilde{v}_2\ll v_1$.)
We regard it as a doubly degenerate $S^1$ with $t \in [0,2\pi]$ and $[2\pi,4\pi]$, implying that the $U(1)_{\rm mix}$ charge of $\zeta$ is two and the $Z_2$ symmetry is unbroken. 
On the other hand, the red curve corresponds to the ${\rm TN}(2,1)$ that is the vacuum manifold at the low energy. We can understand
that it is made by separating two degenerate $S^1$'s so that they lie on opposite points of the poloidal $S^1$. This is nothing but the  spontaneous $Z_2$ breaking at the scale $\tilde v_2$.

Since a torus knot ${\rm TN}(2,1)$ is homotopic to an $S^1$, we have
a nontrivial fundamental group
\begin{eqnarray}
    \pi_1({\rm TN}(2,1)) = \mathbb{Z}\,,
    \label{eq:pi1_TN(2,1)}
\end{eqnarray}
which ensures the existence of global strings.
However, the fact that the minimal loop covering TN(2,1) winds around the $\alpha$-cycle twice and around the $\beta$-cycle once endows interesting properties to the strings as we will see in the next section.

We define nontopological auxiliary quantities as 
\begin{eqnarray}
    \tilde{n}_\alpha(C) &=& \frac{1}{2\pi} \int_C d\alpha\, \\
    \tilde{n}_\beta (C) &=& \frac{1}{2\pi} \int_C d\beta\, 
\end{eqnarray}
where  $C$ is a closed curve or a section of a closed curve
on the physical space on which the fields do not need to lie on the vacuum.
When the fields lie on the vacuum, the quantities need to satisfy 
\begin{eqnarray}
    w = \frac{n_\alpha}{2} = n_\beta \in \mathbb{Z} \ ,
\end{eqnarray}
from Eq.~\eqref{eq:vac_rel}, where $n_\alpha:=\tilde{n}_\alpha(C_{\rm vac})$, $n_\beta:=\tilde{n}_\beta(C_{\rm vac})$,  and $w$ is found to be the winding number for $\pi_1({\rm TN}(2,1))$.

\section{Defects of our model}
\label{sec:defects of our model}

\subsection{Two regimes}
\label{sec:IV.a}

The kinds and length scales of the defects after $\phi$-condensation are different in two regimes divided by 
\begin{align}
 m = m_\ast \ ,%
 \label{eq:saturate_m}
\end{align}
where $m_\ast $ is defined at Eq.~\eqref{eq:two_vacua}.
As we will see below, the origin of the difference comes from the structure of the potential, Eq.~\eqref{eq:V_mod},
\begin{eqnarray*}
V \simeq \!\frac{\lambda_2 v_2^4}{4}\left[\!\!
\left(\left|\frac{\tilde\phi}{v_2}\right|^2\!\!\! - 1\right)^2\!\!\!\!
+ \frac{m}{m_*}\!\!\left\{\left(\frac{\tilde\phi}{v_2}\right)^2\!\!\! + \left(\frac{\tilde\phi^*}{v_2}\right)^2\right\}\!\!
\right]\ .%
\end{eqnarray*}

\subsubsection{Weakly interacting regime}
The potential barrier around the vacuum manifold depends on
the ratio $m/m_\ast$. Let us first explain
the case of
\begin{eqnarray}
    m < m_* \,\quad \left(\tilde v_2 \simeq v_2\right)\,.
    \label{eq:weak_m}
\end{eqnarray}
We call this regime the {\it weakly interacting regime}.
The first two terms in Eq.~\eqref{eq:V} dominate the potential, and we approximately have $|\zeta| = v_1$ and $|\phi| = \tilde{v}_2\simeq v_2$. This means that the torus $\mathcal{T}$ lies deep in the bottom of the high-potential barrier.
Namely, the torus is an approximate vacuum manifold.

\begin{figure}
    \centering
    \includegraphics[width=.85 \linewidth]{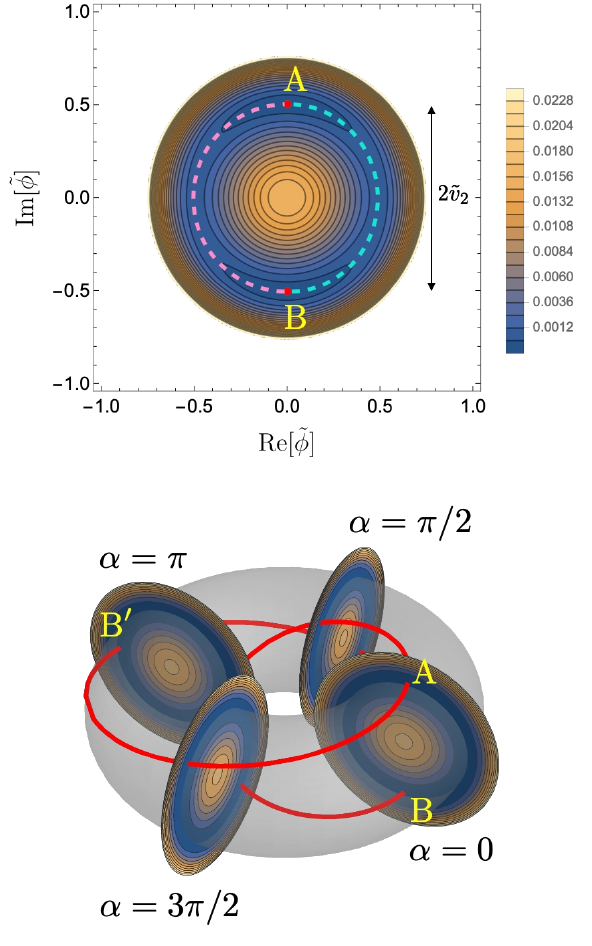}
    \caption{[Top panel] Potential energy on $\tilde{\phi}$-complex plane with $m=0.02 m_*$ in the weakly interacting regime.
    Two minima are located at $\beta-\alpha/2=\pi/2$ (Vacuum A) and $-\pi/2$ (Vacuum B). The origin is the local maximum and two saddle points are located at $\beta-\alpha/2=0$ and $\pi$.  [Bottom panel] Potential energy in $XYZ$-space. 
    The red curve is the vacuum manifold TN(2,1). The gray torus is $\mathcal{T}$. The cross sections are for different $\alpha$: $0$, $\pi/2$, $\pi$ and $3\pi/2$. The red curve represents the vacuum manifold. The gray torus represents the torus~$\mathcal{T}$. The cross section for $\alpha=0$ coincides with the top panel and that for $\alpha=\pi$ rotates around the torus core by $\pi/2$.
    All the dimensional quantities are normalized by $v_1$.
    The parameters are chosen as $(\lambda_1,\lambda_2,v_1,v_2,m)=(1,1,1,1/2,1/400)$. 
    }
    \label{fig:pot_torus_2_weak}
\end{figure}

Figure~\ref{fig:pot_torus_2_weak} shows what the scalar potential looks like. The top panel of Fig.~\ref{fig:pot_torus_2_weak} is the contour plot of %
$V(\tilde\phi)$, which is independent of $\alpha$.
The vacua are represented by the two red dots, $A$ and $B$. 
The potential along the angular direction of $\phi$ with the radius $\tilde{v}_2$ (the cyan and pink dashed curves) remains close to being flat. More precisely, the curve with the quasilowest potential energy on $\alpha$-constant surface which is given by the radial minimum,
\begin{align}
    |\phi|=v_2\sqrt{1-\frac{m}{m_*}\cos(2\beta-\alpha)} \ , 
    \label{eq:weak_ellipse}
\end{align}
i.e., it is not a true circle but an ellipse.~\footnote{Strictly speaking, the approximate manifold is not $\mathcal{T}$ but a slightly deformed torus with the  major radius of $|\zeta|=v_1$ and the minor radius of  $|\phi|$ defined by Eq.~\eqref{eq:weak_ellipse}, depending on $\alpha$ and $\beta$. Since this deformed torus cannot be defined for the strongly interacting regime and it is nearly the same with $\mathcal{T}$ for weakly interacting regime because $\tilde{v}_2\simeq v_2$, we show and refer to $\mathcal{T}$ instead of the deformed torus even for weakly interacting regime for simplicity. }
The bottom panel shows the global structure of the scalar potential. 
$\alpha$-constant surface of $V(\phi)$ looks $V(\tilde{\phi})$ rotated by $\alpha/2$ around the origin.
Namely, when we rotate by $\alpha$ around the $Z$-axis in $XYZ$-space, the potential rotates by $\alpha/2$ along the poloidal $S^1$. Therefore, the potential rotates only by $\pi$ when we go around the $Z$-axis once.
Nevertheless, the potential smoothly connected at $\alpha=0$ since $V(\tilde{\phi})$ is symmetric under the rotation by $\pi$ around the origin, i.e., $V(\phi)=V(\phi e^{-i\pi})=V(-\phi)$.

\begin{figure}
    \centering
    \includegraphics[width=.85 \linewidth]{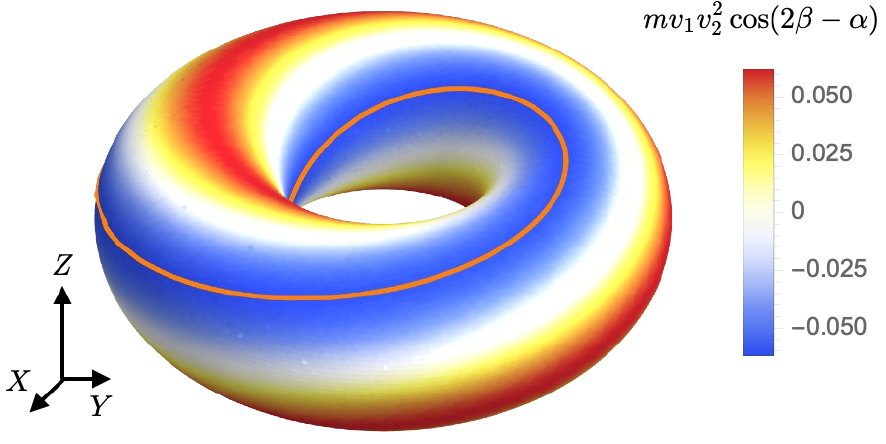}
    \caption{The approximate effective potential energy $mv_1v_2^2\cos(2\beta-\alpha)$ on the torus $\mathcal{T}$, defined at Eq.~\eqref{eq:torus_def}. Blue/red color indicates low/high potential energy. The orange curve corresponds to the vacuum manifold~${\rm TN}(2,1)$. The parameter combination is $(\lambda_1,\lambda_2,v_1,v_2,m)=(1,1,1,1/2,1/400)$. }
    \label{fig:pot_torus}
\end{figure}

The mass spectra consist of two heavy modes corresponding to those changing $|\zeta|$ and $|\phi|$, one exactly massless mode corresponding to the flat direction along ${\rm TN}(2,1)$, and
one light mode [quasi-Nambu-Goldstone (NG) mode] corresponding to the direction orthogonal to ${\rm TN}(2,1)$ and tangential to the torus.
 Figure~\ref{fig:pot_torus} shows the effective potential on the torus by a color density plot.

\subsubsection{Strongly interacting regime}
We call the regime with
\begin{eqnarray}
    m> m_* %
\end{eqnarray}
the {\it strongly interacting regime}, where the third term in Eq.~\eqref{eq:V} is not so small.
As a result, the potential barrier on the torus $\mathcal{T}$ is higher than the potential energy at the torus core 
and we can no longer regard the torus as an approximate vacuum manifold. %
\begin{figure}
    \centering
    \includegraphics[width=.85 \linewidth]{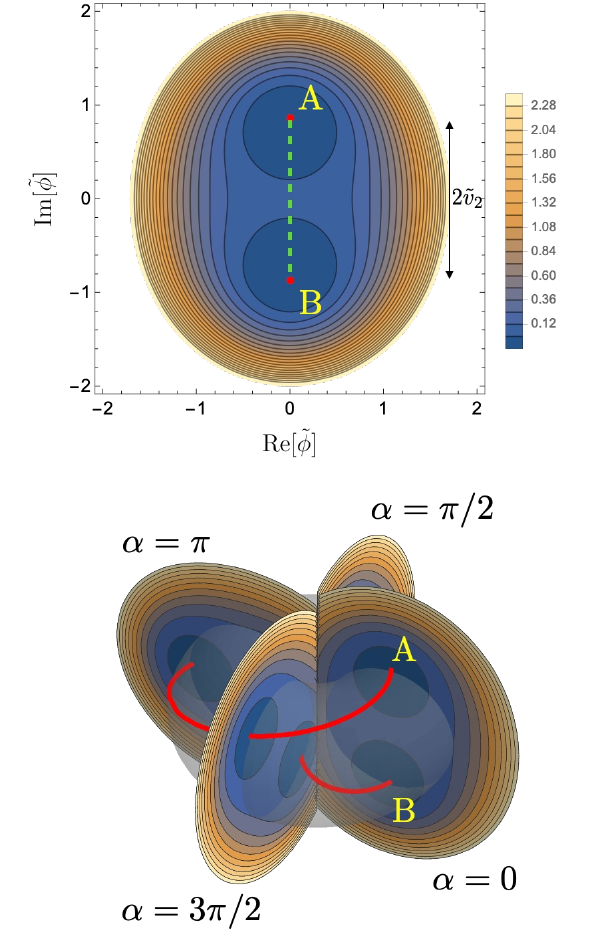}
    \caption{[Top panel] Potential energy on $\tilde{\phi}$-complex plane with $m=2m_*$
    in the strongly interacting regime.
    Two minima are located at $\beta-\alpha/2=\pi/2$ (Vacuum A) and $-\pi/2$ (Vacuum B) and a saddle point is located at the origin. 
    [Bottom panel] Potential energy in $XYZ$-space. 
    See the caption of Fig.~\ref{fig:pot_torus_2_weak}.
    All the dimensional quantities are normalized by $v_1$.     The cross sections for different $\alpha$ overlap each other at $Z$-axis since the map defined by \eqref{eq:XYZ} into $XYZ$-space is only well-defined within $|\phi|<v_1$. The parameters are chosen as $(\lambda_1,\lambda_2,v_1,v_2,m)=(1,1,1,1/2,1/4)$.}
    \label{fig:pot_torus_2_strong}
\end{figure}
Figure~\ref{fig:pot_torus_2_strong} shows the scalar potential in the strongly interacting regime. The top panel of Fig.~\ref{fig:pot_torus_2_strong} is the contour plot of %
$V(\tilde\phi)$.
The global structure is similar to that in the weakly interacting regime as shown in the bottom panel of Fig.~\ref{fig:pot_torus_2_strong}.
However, the poloidal $S^1$ structure no longer exists.

\subsection{Mass spectrum}
We examine the mass spectrum appearing in this model. There are four states, and they have different spectrum in the two regimes.  Here we show the typical values under the assumption of the energy scales examplified in Sec.~\ref{sec:the model}. The full expression is shown in Appendix~\ref{app:mass}.

For weakly interacting regime $\tilde{m}=m/m_\ast <1$,  we find a massless NG boson and a quasi-NG boson with $m_{\rm qNG}=2\sqrt{m v_1}$ from the linear combination of $\arg \zeta$ and $\arg \phi$. In addition, there exist two massive states with $m_{\rm 1}=\sqrt{2\lambda_1}v_1$ and $m_{\rm 2}=\sqrt{2\lambda_2}v_2$ from the linear combination of $|\zeta|$ and $|\phi|$.
Concretely, 
\begin{align}
    m_\ast = 5.0 \times 10^{-4}\Bigl(\frac{10^{16}{\rm GeV}}{v_1}\Bigr)\Bigl(\frac{v_2}{10^2 {\rm GeV}}\Bigr)^2 \lambda_2 \, {\rm eV} \ ,
\end{align}
and the masses are evaluated as
\begin{align}
    m_{\rm q NG}& = 1.4 \times 10^2 \,\tilde{m}^{1/2}\Bigl(\frac{v_1}{10^{16} {\rm GeV}}\Bigr)^{1/2} {\rm GeV}\nonumber \\
    &<1.4 \times 10^2 \Bigl(\frac{v_1}{10^{16} {\rm GeV}}\Bigr)^{1/2} {\rm GeV} , 
    \label{eq:qNG}
\end{align}
and
\begin{align}
    m_1=1.4\times 10^{16} {\rm GeV}\sim v_1\ , \quad
    m_2=1.4\times 10^{2} {\rm GeV} \sim v_2\ .
    \label{eq:m1m2}
\end{align}

For strongly interacting regime $\tilde{m}>1$, there exist one NG boson and three massive modes. The first radial massive mode has the  mass $m_1=\sqrt{2\lambda_1}$ (\eqref{eq:m1m2}) as well. The radial second massive mode $m_2$ and angular massive mode $m_3$ has the mass $2\sqrt{m v_1}$ (the expression in \eqref{eq:qNG}). These masses satisfy
\begin{align}
    m_2=m_{3} > 1.4 \times 10^2 \Bigl(\frac{v_1}{10^{16} {\rm GeV}}\Bigr)^{1/2} {\rm GeV}  \ .
\end{align}

\subsection{Defects}

\subsubsection{Topological and nontopological $\zeta$-strings}
\label{sec:zeta_string}

In high energies, the vacuum manifold is just $S^1$ in $\alpha$-direction, and therefore, there exist 
strings winding around $\alpha$-cycle, which we call $\zeta$-strings. 
These all are stable in high energy above $\tilde{v}_2$. As seen in the previous section, the vacuum manifold in low energies is TN(2,1), and thus, $\zeta$-strings with even $n_\alpha$ can smoothly wrap around the vacuum. Therefore, such $\zeta$-strings are topological and stable even in low energies. However, the SSB at $\tilde{v}_2$ destabilizes $\zeta$-strings with odd $n_\alpha$. 
From now on, we focus on the $\zeta$-strings with $n_\alpha=\pm 1$, which are important both cosmologically and numerically. 
When we go on the vacuum by $2 \pi$ for $\alpha$-cycle, the initial point and the end point are different by $\pi$ for $\beta$-cycle. 
Therefore, one has to escape once from ${\rm TN}(2,1)$ and come back to ${\rm TN}(2,1)$. %
To minimize the gradient energy cost, the escape is forced to be in the range with a slight change in the $\alpha$-direction, namely, it is constrained on approximately $\alpha$-constant surface.
The behaviour during leaving from ${\rm TN}(2,1)$ is different for the weakly interacting regime and for the strongly interacting regime. For the former, the configuration goes around the $\beta$-cycle by $\pm \pi$ to avoid the torus core with high energy. The approximate path is shown by the cyan/pink dashed curve in Fig.~\ref{fig:pot_torus_2_weak}. For the latter, it goes near
the torus core, where the energy density is lower than that of the torus surface.
The approximate path is shown by the green dashed curve in Fig.~\ref{fig:pot_torus_2_strong}. On the approximately $\alpha$-constant surface, the initial point and the end point on TN(2,1) look two discrete vacua 
as a result of the spontaneous breaking of $Z_2$. (See the top panels of Figs.~\ref{fig:pot_torus_2_weak} and \ref{fig:pot_torus_2_strong}.) Therefore, the escape from TN(2,1) for $\zeta$-strings corresponds to a domain-wall solution connecting these two vacua. Namely, each $\zeta$-string with $n_\alpha=\pm 1$ is inevitably attached to a wall.~\footnote{Each $\zeta$-string with an odd winding number $|n_\alpha| > 1$ is also attached to a wall from the same reason.}

Since the meaning of $n_\alpha=\pm 1$ is not obvious in the previous paragraph, we give a more precise definition of the winding of $\zeta$-strings along $\alpha$-cycle. Let us consider a contour surrounding a $\zeta$-string from far away in the physical space, $C^{\zeta}$. In high energies, the fields can be in the vacuum at any points on $C^{\zeta}$, and thus, $n_\alpha=\tilde{n}_\alpha(C^\zeta)=\pm 1$. However, as we mentioned, in low energies, all the points on $C^{\zeta}$ cannot be in the vacuum. 
Fields on a section of $C^\zeta$ need to be away from the vacuum and the other part of $C^\zeta$ can be in the vacuum. We name them $C^\zeta_{\rm dw}$ and $C^\zeta_{\rm vac}$, respectively. $C^\zeta=C^\zeta_{\rm dw}\cup C^\zeta_{\rm vac}$. Now, $\tilde{n}_\alpha(C^\zeta)$ is not a topological value, but is still an integer number, $\tilde{n}_\alpha(C^\zeta)=\pm 1$, because of the single valuedness of $\zeta$-field. In low energies, though $n_\alpha$ cannot be defined as a topological number for an isolated $\zeta$-string with $\tilde{n}_\alpha(C^\zeta)=\pm 1$, we refer to $n_\alpha$ instead of $\tilde{n}_\alpha(C^\zeta)$ for simplicity. 
As we will see in Sec.~\ref{sec:string-walls}, a pair of them is topological. 

\subsubsection{Nontopological domain walls}
\label{sec:walls}

As previously mentioned, an approximately $\alpha$-constant surface of the effective potential has two discrete vacua as an outcome of the spontaneous breaking of $Z_2$, and there exist domain-wall solutions connecting these two vacua.

For the weakly interacting regime,
a wall solution corresponds to one of the orbits connecting two vacua along a half of $S^1$ of $\beta$-cycle, i.e., near the surface of the torus $\mathcal{T}$. (The orbits are shown in Fig.~\ref{fig:pot_torus_2_weak} by the cyan and pink dashed curves.)
We evaluate the width and the tension as (see Appendix~\ref{app:field_config} for the derivation),
\begin{align}
    \delta^{\rm (weak)}_{\rm DW}    \sim (2\sqrt{|m| v_1})^{-1}   \ ,
\label{eq:dw_width_weak}
\end{align}
and
\begin{align}
    \sigma^{\rm (weak)}= 4\sqrt{m v_1} v_2^2 \ .
        \label{eq:sigma_weak}
\end{align}
Since the domain-wall solution is described by $\arg \tilde{\phi}=\beta-\alpha/2$, 
the quantity $\beta-\alpha/2$ or its twice $2\beta-\alpha$ is useful to identify walls in numerical simulation.
For the strongly interacting regime,
a domain-wall solution corresponds to the orbits connecting two vacua through the torus core. (The orbits are shown in Fig.~\ref{fig:pot_torus_2_strong} by the green dashed curve.)
As calculated in Appendix~\ref{app:field_config}, we find the domain-wall solution in the infinite plane limit (without a string boundary) and evaluate the wall width as
\begin{align}
    \delta^{\rm (strong)}_{\rm DW}\sim(\sqrt{\lambda_2} \tilde{v}_2)^{-1}   \ ,
\label{eq:dw_width_strong}
\end{align}
and the wall tension as
\begin{align}
    \sigma^{\rm (strong)}
    = \frac{2 \sqrt{2}}{3} \sqrt{\lambda_2} \tilde{v}_2^3 \ .
    \label{eq:sigma_strong}
\end{align}
Since the wall solution is described by $\Im \tilde{\phi}$ and $\Re \tilde{\phi}=0$, the quantity $|\tilde{\phi}|=|\phi|$ is useful to identify domain walls in numerical simulation.
By comparing these quantities Eqs.~\eqref{eq:dw_width_weak}-\eqref{eq:sigma_strong}, we can see that
the domain walls are less massive and broader for the weakly interacting regime, which will be confirmed numerically in the next section.

The walls are not topologically stable since the orbits connecting the vacua A and B can be continuously contracted to a single point through the vacuum manifold.
Nevertheless, we expect that the wall is sufficiently metastable since an intermediate state (for example a curve connecting ${\rm A}'$  and B  in Fig.~\ref{fig:pot_torus_2_weak}) 
clearly needs a large gradient-energy cost.

For the weakly interacting regime, it looks that there are two different wall solutions corresponding to the cyan and pink dashed curves in the top panel of Fig.~\ref{fig:pot_torus_2_weak}. However, these are topologically indistinguishable. We can move a cyan dashed curve by $\pi$ along $\alpha$-cycle with no energy cost, and, after we move it, it lies on the pink dashed curve. When we map a path with a direction crossing a wall on the physical space to the field space, we can specify the direction on the field space, i.e., if $\arg \tilde{\phi}=\beta-\alpha/2$ is increasing (${\rm B} \to {\rm A}$ for the cyan dashed curve) or decreasing (${\rm A} \to {\rm B}$ for the cyan dashed curve). The direction does not change even if we move the cyan/pink dashed curve by $\pi$ along $\alpha$-cycle. Therefore, there are two kinds of walls, one of which has an increasing phase ($\tilde{n}_\beta
>0$) and the other has a decreasing phase ($\tilde{n}_\beta
<0$).  Even for the strongly interacting regime, the orbit does not straightly go through the torus core but it just passes near the torus core because of the gradient energy cost and $\beta$-phase advances not discontinuously but continuously, i.e., it earns a finite $\tilde{n}_\beta
$.

To reduce the energy cost, one of the kinds of the walls is chosen when a $\zeta$-string is attached to a wall. A $\zeta$-string with $n_\alpha=1$ is attached to a wall with $\tilde{n}_\beta(C_{\rm dw}^\zeta)=-1/2$, since $\tilde{n}_\beta(C_{\rm vac}^\zeta)=\tilde{n}_\alpha(C_{\rm vac}^\zeta)/2=1/2$. 
Similarly, one with $n_\alpha=-1$ is attached to a wall with $\tilde{n}_\beta(C_{\rm dw}^\zeta)=1/2$. (See Fig.~\ref{fig:weak_phase}.) For the weakly interacting regime, the other choices make a loss in the potential energy since the orbit surrounds the torus core.  For the strongly interacting regime, the other choices make a loss in the gradient energy. 
We note that the walls are not only terminated by the $\zeta$-strings but also form closed walls.

\subsubsection{Nontopological $\phi$-string}
\label{phi_strings}

For the weakly interacting regime, there exist a different type of strings from $\zeta$-strings, strings winding around $\beta$-cycle, which we call $\phi$-strings, in low energies below $\tilde{v}_2$.~\footnote{$\phi$-strings have no corresponding defects in high energies.}
This is because the torus core has high potential energy~$\sim \lambda_2 v_2^4$ and 
the bumps %
on the torus $\mathcal{T}$, which has the potential energy maximal on the torus surface,
\begin{align}
  |\phi|=v_2\sqrt{1-\frac{m}{m_*}}\simeq v_2
\ , \qquad 
    2\beta -\alpha= 0  \mod 2\pi \ ,
    \label{eq:bump}
\end{align}
carry a lower energy~$\sim m v_1 v_2^2$. 
To reduce the gradient energy, $\phi$-strings wind around $\beta$-cycle on approximately $\alpha$-constant surface of the torus. Again, we focus on the strings with the minimal winding, $\tilde{n}_\beta(C^\phi)=\pm 1$, where $C^\phi$ is a contour surrounding a $\phi$-string. The $\phi$-stings climb over the bumps twice. We can divide the contour into two pieces, e.g., for a $\phi$-string with $\tilde{n}_\beta(C^\phi)=1$, a path is from vacuum A to vacuum B and the other is from B to A, (pink and cyan dashed curves in Fig.~\ref{fig:pot_torus_2_weak}, respectively.) These are nothing but nontopological domain walls found in Sec.~\ref{sec:walls}.~\footnote{The $\phi$-string can be regarded as a daughter wall inside a mother wall associated with the sequential SSB $Z_2 \times \mathscr{C}' \to \mathscr{C}' \to I$ where the second SSB takes place only inside the mother wall~\cite{Eto:2023gfn}.}
Therefore, $\phi$-strings are inevitably attached to two domain walls.
One can see each $\phi$-string is attached to the same kind of the domain walls; for $\tilde{n}_\beta(C^\phi)=1$, both of the domain walls attached carry $\tilde{n}_\beta(C^\phi_{\rm dw })=1/2$, and for $\tilde{n}_\beta(C^\phi)=-1$, both carry $\tilde{n}_\beta(C^\phi_{\rm dw })=-1/2$.

We summarize the values of $\tilde{n}_\alpha$ and $\tilde{n}_\beta$ for each nontopological defect in Table~\ref{tab:chages}.

\begin{table}[htbp]
\centering
\begin{tabular}{c|c|c|c|c|c|c}
       & \multicolumn{2}{c|}{$C$} & \multicolumn{2}{c|}{$C_{\rm vac}$} & \multicolumn{2}{c}{$C_{\rm dw}$}\\ 
       & $\tilde{n}_\alpha$ & $\tilde{n}_\beta$& $\tilde{n}_\alpha$ & $\tilde{n}_\beta$& $\tilde{n}_\alpha$ & $\tilde{n}_\beta$ \\ \hline\hline
      $\zeta$-string & $\pm 1$ & 0 & $\pm 1$ & $\pm \frac{1}{2}$ & 0 & $\mp \frac{1}{2}$ \\ \hline
      wall &  &   &   &   & 0 & $\pm  \frac{1}{2} $\\ \hline
      $\phi$-string & 0 & $\pm 1$ & 0  & 0 & 0 & $\pm \frac{1}{2} \times 2$\\ \hline
\end{tabular}
\caption{A list of $\tilde{n}_\alpha$ and $\tilde{n}_\beta$ for $\zeta$-strings, walls, $\phi$-strings. $C$ is a closed contour surrounding each defect. $C_{\rm vac}$  is the sum of sections of $C$ where the fields are in vacuum, and $C_{\rm dw}$ is the sum of the rest sections of $C$ where the fields are away from vacuum and the path crosses a wall. $C=C_{\rm vac}\cup C_{\rm dw}$. $\tilde{n}_\beta(C_{\rm dw})$ for $\phi$-strings is doubled because each $\phi$-string is accompanied by two walls.}
\label{tab:chages}
\end{table}

\subsubsection{Topological string-walls}
\label{sec:string-walls}

\begin{figure}[h]
    \centering
    \includegraphics[width=1.0 \linewidth]{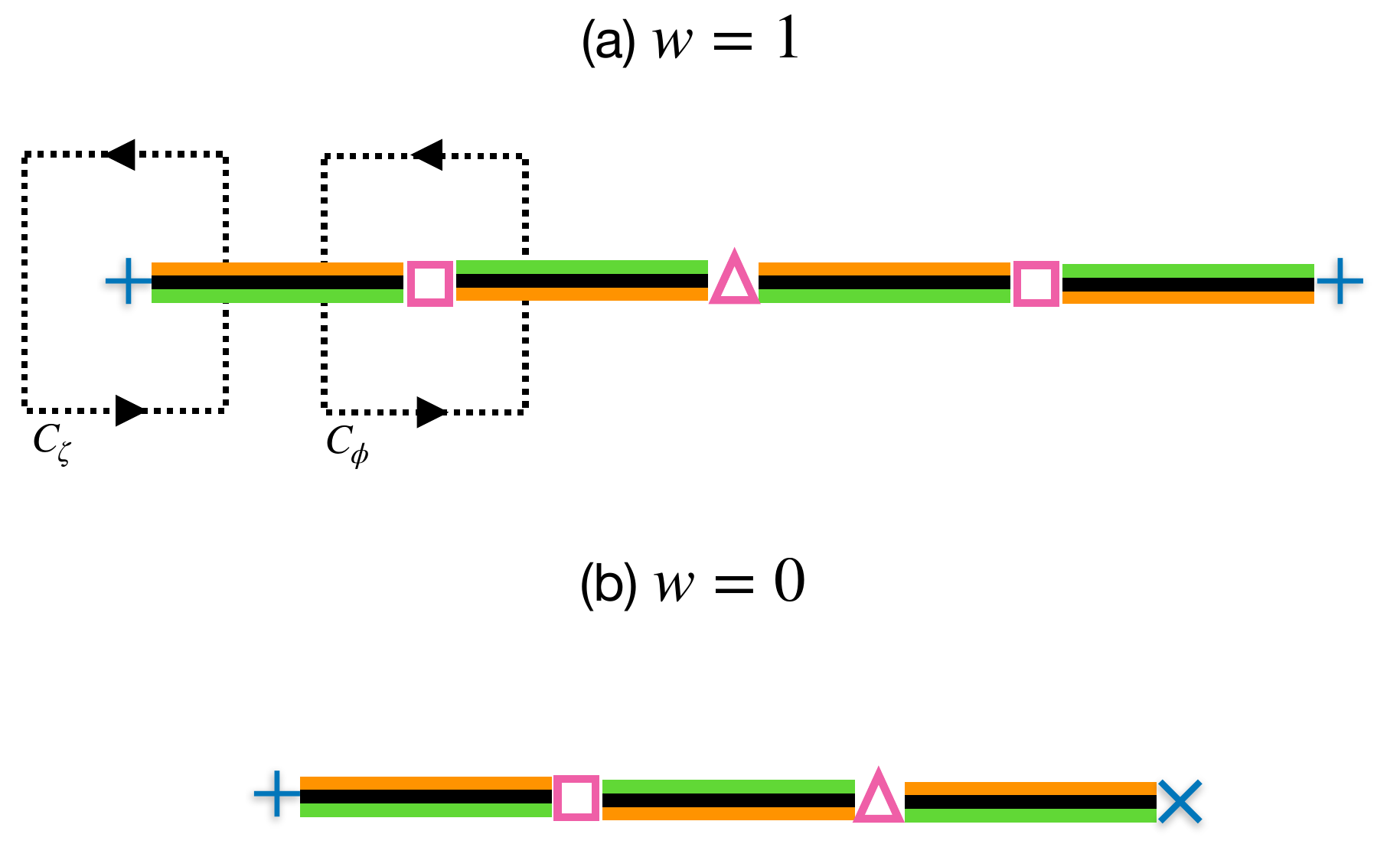}
    \caption{(a)/(b) schematically describes the configuration of a $w=1$/$w=0$ string-wall for the weakly interacting regime. Blue plus/cross mark represents $\zeta$-string with $n_\alpha=1$/$-1$, %
    respectively. Red rectangle/triangle mark represents $\phi$-string with $\tilde{n}_\beta=-1/1$,
    respectively. The colors show the value of $2\beta-\alpha$: the white color shows $\pi \mod 2\pi$ corresponding to the vacuum, the black color shows $0\mod 2\pi$, corresponding to the center of the wall, i.e., the top of the bump on the torus, the green color shows $\pi/2 \mod 2\pi$, and the orange color shows $-\pi/2 \mod 2\pi$. We can see the direction of the phase from the order of these colors. (The color choices are the same as those used in Fig.~\ref{fig:TE_weak}.) 
    For the strongly interacting regime, $\phi$-strings are not confined and are extended along the wall.}
    \label{fig:weak_phase}
\end{figure}

To make $\zeta$-strings topological in low energies, they need to be paired with each other. $\zeta$-strings can be paired by attaching the ends of the walls emanating from the $\zeta$-strings. 
Obviously, $\zeta$-strings with the opposite $\tilde{n}_\alpha(C)$ can be paired, which unwinds around the vacuum manifold~TN(2,1). Such a pair is topologically trivial, which does not carry a winding number defined by $U(1)_{\rm mix}$, $w=0$. In addition, $\zeta$-strings with the same $\tilde{n}_\alpha(C)$ can also be paired, which winds around TN(2,1) once.  Such a pair is topologically nontrivial, which carries $w=\pm 1$. 
A pair of $\zeta$-strings with $\tilde{n}_\alpha(C)=1$ earn $\tilde{n}_\beta(C_{\rm vac})=1/2\times 2$, resulting in $n_\beta=1$ for a closed curve surrounding both of the strings in the physical space. This means the existence of a $\phi$-string on the wall connecting the $\zeta$-strings for the weakly interacting regime.
Namely, two $\zeta$-strings with $n_\alpha=1(-1)$ connected by a $\phi$-string with $\tilde{n}_\beta=1(-1)$ is topological. In addition, any number of a composite of $\phi$-strings with opposite $\tilde{n}_\beta$ (connected by walls with each other) can be inserted on the wall connecting the $\zeta$-strings since it does not change $w$. From the observation of $\tilde{n}_\beta$, when it is embedded in the wall, a $\phi$-string with $\tilde{n}_\beta=1$ needs to be the neighborhood of $\zeta$-string with $n_\alpha=1$ and/or $\phi$-string with $\tilde{n}_\beta=-1$, and vice versa. (See Fig.~\ref{fig:weak_phase}.) 
For the strongly interacting regime, $\phi$-strings do not exist and the $\zeta$-strings are connected just by the walls. In other words, a  wall can be understood as an extended object of a $\phi$-string by the strong interaction between the fields.

\begin{figure}
    \centering
        \vspace*{-5mm}
    \centering
        \includegraphics[width=1. \linewidth, keepaspectratio]{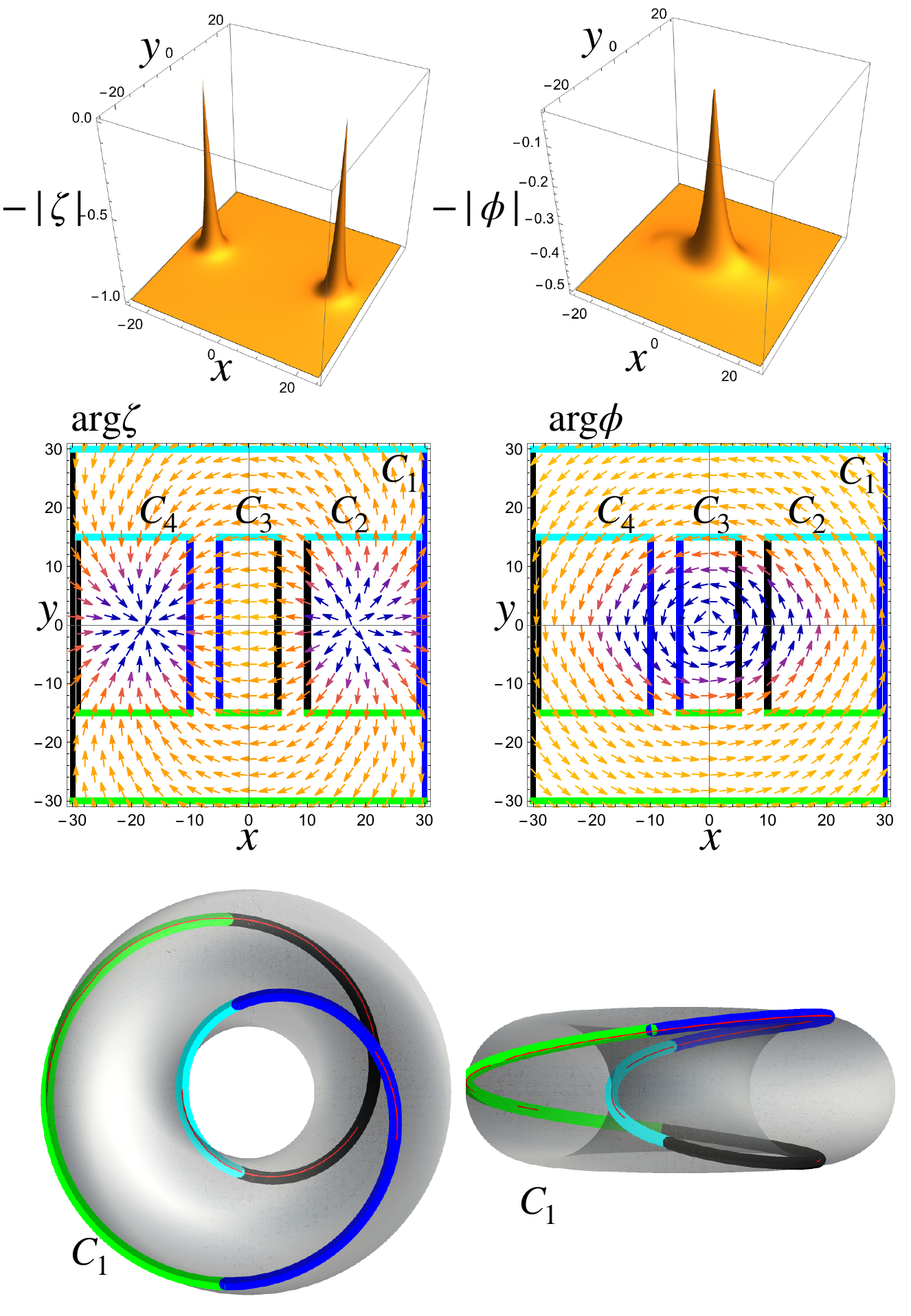}
    \caption{A nonequilibrium $w=1$ string-wall numerically found for  $\lambda_1=\lambda_2=1$, $v_2=1/2$, and $m=1/400$ in $v_1$ unit, which is in the weak interaction regime. The top-left (-right) panel show $-|\zeta|$ ($-|\phi|$) near the string-wall.
    We see that the $\phi$-string is confined.
    In the middle panels, we show the phases of $\zeta$ and $\phi$ by the vector plots of $({\rm Re}\phi,{\rm Im}\phi)$. The contours $C_2$ and $C_4$ surround one of the $\zeta$-strings at the end of the walls. $C_3$ surrounds the $\phi$ string connecting two $\zeta$-strings through the wall. $C_4$ surrounds all the ingredients of the $w=1$ string-wall. The bottom panels show the image of the contour $C_1$ into $XYZ$-space, showing the field values, $(\alpha, |\phi|, \beta)$. We find that the image of $C_1$ covers the torus knot vacuum manifold (shown by a red curve) once, supporting that the string-wall is topological and $w=1$. The gray torus is $\mathcal{T}$. 
    The gray tone indicates the potential value at each point on the torus: the black/white color shows the lowest/highest energy. The bottom-left (-right) is from the top (front) view of the torus.
    }
        \label{fig:torus_weak}
\end{figure}

\begin{figure}
    \centering
        \vspace*{-5mm}
    \centering
        \includegraphics[width=1. \linewidth, keepaspectratio]{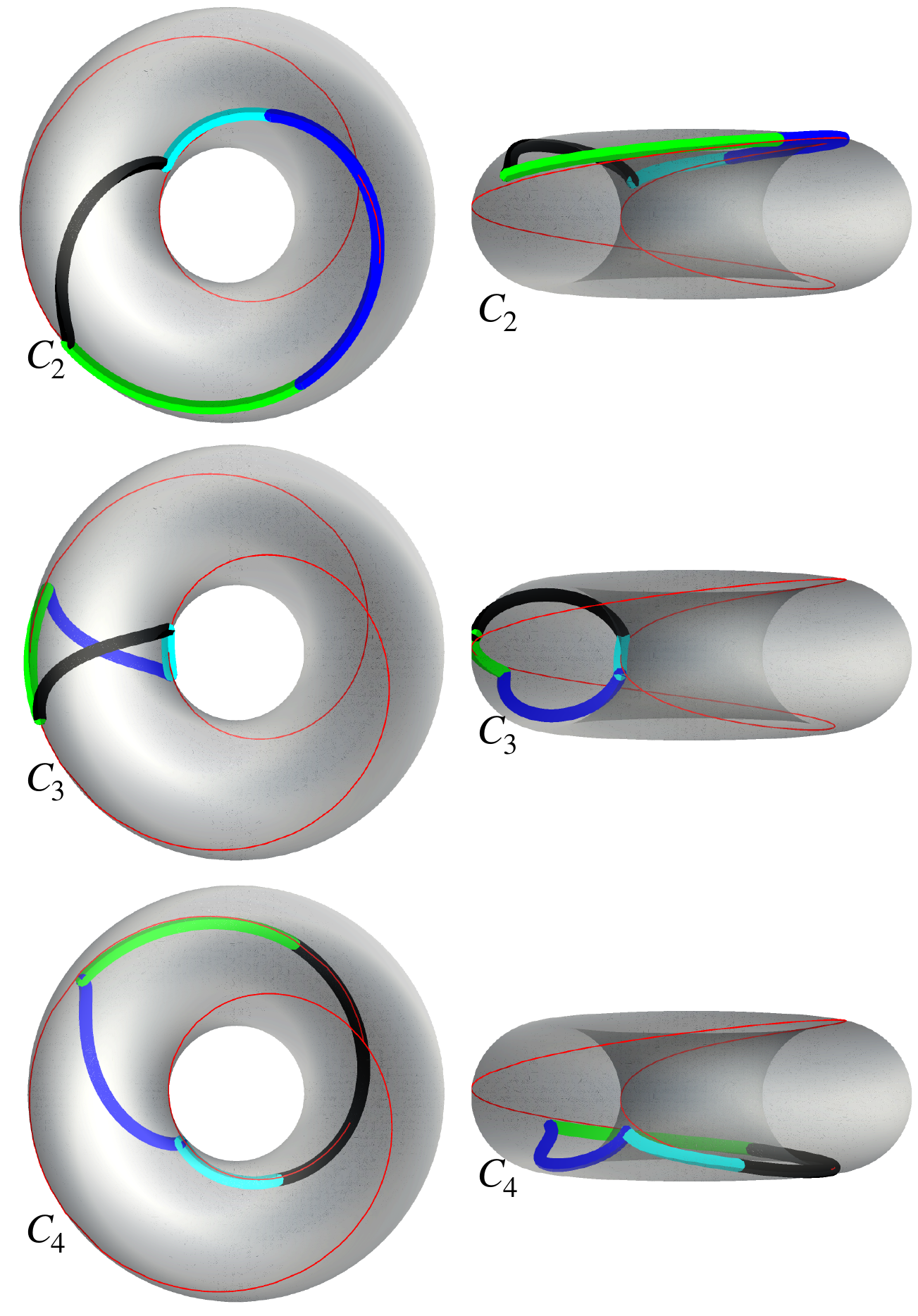}
    \caption{The top/middle/bottom panels  show the image of the contour $C_2$/$C_3$/$C_4$ in Fig.~\ref{fig:torus_weak} into $XYZ$-space. 
    See the caption of Fig.~\ref{fig:torus_weak} for the detailed explanation.
    In the right panels, we can see that the fields pass near the surface of the torus $\mathcal{T}$ for the walls. 
    The sum of the images of $C_2$, $C_3$ and $C_4$ approximately coincides with the image of $C_4$, shown in the bottom panels of Fig.~\ref{fig:torus_weak}.  } %
        \label{fig:torus_weak2}
\end{figure}

We perform a numerical simulation of a string-wall for the weakly interacting regime using a standard relaxation method. A $w=1$ string-wall is stable and seems to have a configuration where the repulsive force between $\zeta$-strings and the attractive force of the walls balances. However, the distance between the strings of the balance is  large because the wall tension is weak.
In Figs.~\ref{fig:torus_weak} and \ref{fig:torus_weak2}, we show a $w=1$ string-wall with one $\phi$-string. The $\zeta$-strings go away from each other because the repulsive force wins the attractive force at this distance between the $\zeta$-strings. In Fig.~\ref{fig:torus_weak}, the top panel shows
$\phi$-string is confined and the walls are faint as expected analytically. We also show $\arg \zeta = \alpha$ and $\arg \phi=\beta$ in the middle panels.
In the bottom panel, we map the field values along a contour $C_1$ surrounding the string-wall into $XYZ$-space. We can see that the string-wall wraps around the vacuum manifold once and is topological.

\begin{figure}
    \centering
        \vspace*{-5mm}
    \centering
        \includegraphics[width=1. \linewidth, keepaspectratio]{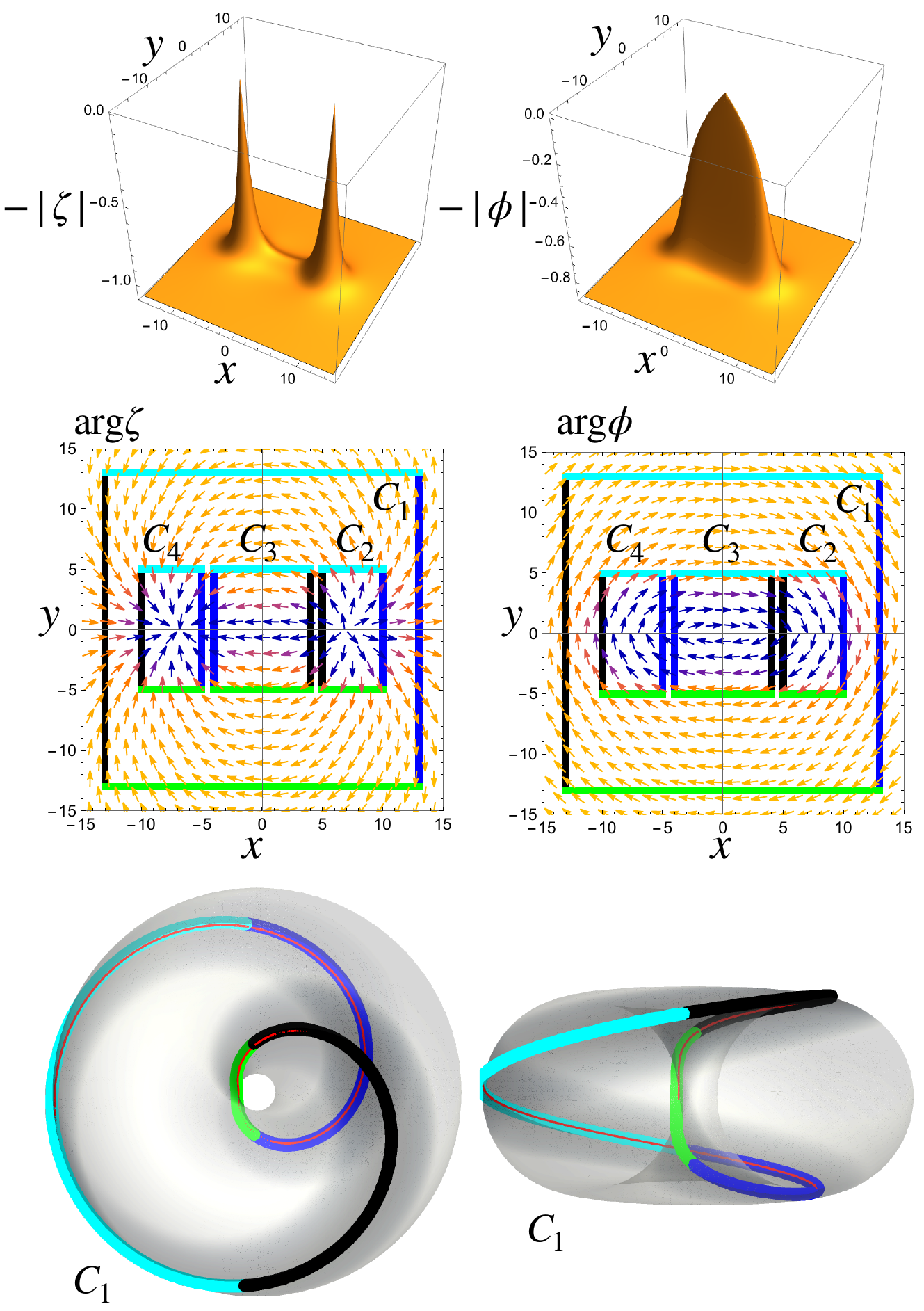}
    \caption{An equilibrium $w=1$ string-wall numerically found for $\lambda_1=\lambda_2=v_1=1$, $v_2=1/2$, and $m=1/4$ in $v_1$ unit,  which is in the strongly interaction regime. See the caption of Fig.~\ref{fig:torus_weak} for the detailed explanation.
     We see that the $\phi$-string is extended between the two $\zeta$-strings. We again find that the image of $C_1$ covers the torus knot vacuum manifold (shown by a red curve) once, supporting that the string-wall is topological and $w=1$. We note that the gray torus is $\mathcal{T}$. %
    }
        \label{fig:torus_strong}
\end{figure}

\begin{figure}
    \centering
        \vspace*{-5mm}
    \centering
        \includegraphics[width=1. \linewidth, keepaspectratio]{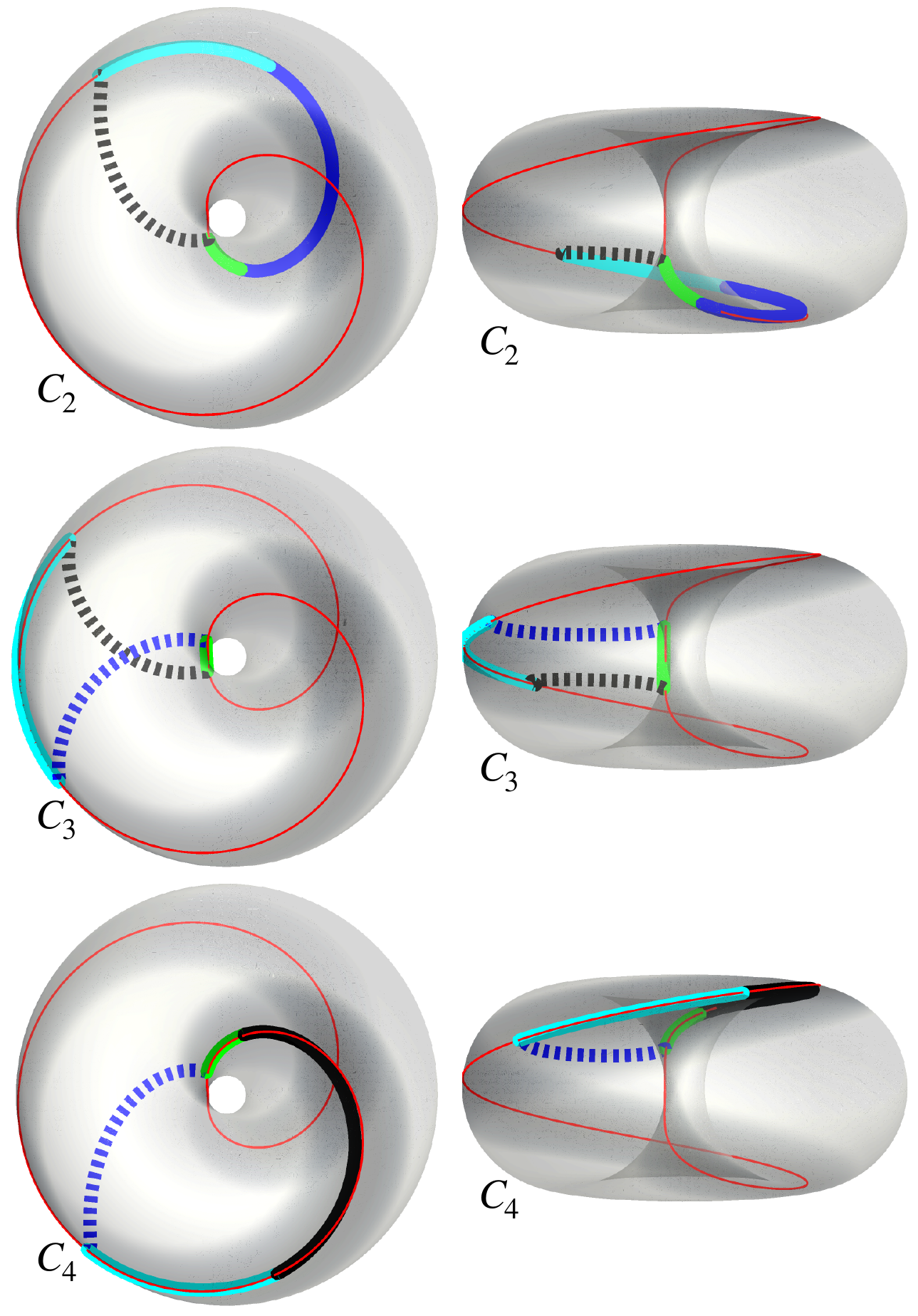}
    \caption{The top/middle/bottom panels  show the image of the contour $C_2$/$C_3$/$C_4$ in Fig.~\ref{fig:torus_strong} into $XYZ$-space. See the caption of Fig.~\ref{fig:torus_weak} for the detailed  explanation. In the right panels, we can see that the fields pass near the torus core for the walls, shown by dashed lines. The sum of the images of $C_2$, $C_3$ and $C_4$ approximately coincides with the image of $C_4$, shown in the bottom panels of Fig.~\ref{fig:torus_strong}.
     }
        \label{fig:torus_strong2}
\end{figure}

We perform a similar numerical simulation for a  string-wall in the strongly interacting regime. A $w=1$ string-wall is topologically stable and has an equilibrium configuration where the repulsive force between $\zeta$-strings and the attractive force from the wall tension becomes equal and balances.
This equilibrium configuration is shown in Figs.~\ref{fig:torus_strong} and \ref{fig:torus_strong2}. The top-right panel of Fig.~\ref{fig:torus_strong} shows that there exists a single point where $|\phi|=0$ is reached as  a result of $n_\beta=1$.

$w=0$ string-walls are unstable and shrink to vanish since they are topologically trivial. 
We revisit these features of string-walls in cosmological simulations in the next section.



We compare the length scales included in string-walls.~\footnote{This estimation is field-value-based and does not correspond to the length scales seen in energy-density-based plots.}
We consider the weakly interacting regime first.
After $\zeta$-condensation, $\zeta$-strings have a normal core size of strings as
\begin{align}
    \delta_{\zeta  } \sim (\sqrt{\lambda_1} v_1)^{-1}  \ .
\label{eq:delta_zeta}
\end{align}
After $\phi$-condensation, 
$\phi$-strings are formed and have a length scale of (See Appendix~\ref{app:field_config})
\begin{align}
    \delta^{\rm (weak)}_{\phi} \sim (\sqrt{\lambda_2} v_2)^{-1}\ ,
\end{align}
which is smaller than the wall thickness~\eqref{eq:dw_width_weak}
Therefore, there is a hierarchy between the length scales for string-walls as
\begin{align}
\delta_{\zeta} <  \delta_{\phi}^{\rm (weak)} < \delta_{\rm DW}^{\rm (weak)}  \ .
\end{align}
In the weak interaction limit, $m\rightarrow 0$, the wall thickness becomes much larger than the string core sizes,
\begin{align}
     \delta_{\phi}^{\rm (weak)}  \ll \delta_{\rm DW}^{\rm (weak)}  \ .
\label{eq:dw_width_weak_limit}
\end{align}
For the strongly interacting regime, $\zeta$-strings have a length scale of \eqref{eq:delta_zeta} and walls have that of \eqref{eq:dw_width_strong}.
Therefore, 
\begin{align}
    \delta_{\zeta} < \delta_{\rm DW}^{\rm (strong)} \ .
\label{eq:strong_width_comp}
\end{align}
If $m \lesssim  v_1$, these length scales become comparable.~\footnote{More precisely, because of the strong interaction between the fields, the backreaction from $\phi$ affects $\zeta$-strings. In that case, the $\zeta$-string core size becomes smaller than estimated by \eqref{eq:delta_zeta} because $|\zeta|=\tilde{v}_1=v_1\sqrt{1+\frac{2m^2}{\lambda_1\lambda_2 v_1^2}}$ and $|\phi|=v_2\sqrt{1+\frac{2m \tilde{v}_1}{\lambda_2 v_2^2}}$. In Figs.~\ref{fig:torus_strong} and \ref{fig:torus_strong2}, $m/v_1\sim 1/4$ is not small and we use the precise values shown here for the vacuum manifold in Figs.~\ref{fig:torus_strong} and \ref{fig:torus_strong2}.}

\section{Cosmological simulation}

To confirm the structure of the defects discussed above, we perform the field-theoretic simulations in the expanding Universe.

For a cosmological setup, we add the finite temperature effect
to the potential, %
\begin{align}
    \Delta V(\zeta, \phi )
     =\frac{\lambda_1 }{6} T^2 |\zeta|^2  
    +\frac{\lambda_2 }{6} T^2 |\phi|^2   \ .
    \label{eq:temp}
\end{align}
The first phase transition occurs at 
\begin{align}
    T_1 = \sqrt{3} v_1  \ ,
\end{align}
with $\langle | \phi | \rangle = 0 $, and
the second phase transition occurs at 
\begin{align}
    T_2 =   \sqrt{3} \tilde{v}_2  \ .
\end{align}
We note that the temperature of the second phase transition depends on the strength of the interaction.
In the expanding Universe, the metric is given as
\begin{align}
    ds^2 = a(t)^2 (-d t^2  + d \mathbf{x}^2).
\end{align}
The equations of motion %
found from \eqref{eq:action} with the finite temperature effect~\eqref{eq:temp} are 
\begin{align}
     &\ddot{\zeta} +2 \mathcal{H} \dot{\zeta} -\delta^{ij} \partial_i\partial_j \zeta \nonumber \\
     &=-a^2 \Bigl( \lambda_1\zeta (|\zeta|^2-v_1^2) +m \phi^2\
     +\frac{\lambda_1}{3} T^2 \zeta\Bigr)\ ,  
     \label{eq:zeta}
\end{align}
and
\begin{align}
     &\ddot{\phi} + 2 \mathcal{H} \dot{\phi}-\delta^{ij} \partial_i\partial_j \phi \nonumber \\
     &=-a^2 \Bigl( \lambda_2 \phi (|\phi|^2-v_2^2) +2 m^\ast \zeta \phi^\ast
     + \frac{\lambda_2}{3}T^2 \phi \Bigr)\ , 
     \label{eq:phi}
\end{align}
where the dot denotes the time derivative in the conformal time and $\mathcal{H}:=\dot{a}/a$ is the conformal Hubble parameter.

We use a two-dimensional comoving box with periodic boundaries for the simulation and impose a thermal initial condition. (for details, see Appendix~\ref{App:Initial}).
We choose the initial time so that the background temperature is $T_{\rm in}=2 T_1$.
We employ the leap-frog method for the time evolution and 
the second-order central finite differences for the spatial derivatives.
To identify $\zeta$-strings and $\phi$-strings, we use the algorithm developed by Ref.~\cite{Yamaguchi:2002sh}.
We choose the following model parameters:
\begin{align}
    & \lambda_1 = \lambda_2 = 1\ ,%
    & v_1 = 3.556 \times 10^{16} {\rm [GeV]} \ , %
\end{align}
and $v_2$ and $m$ are left as control parameters.
We normalize all the dimensional quantities by $v_1$ in the following.
Then, the first phase transition occurs at $t_1: = 3.33$, corresponding to $T_1:= 1.73$.
We summarize the control parameters
in Table~\ref{tab:parameters}.
\begin{table*}[htbp]
\centering
\begin{tabular}{c||c|c||c|c||c|c|c||c|c}
      Run & $v_2$ & $m/m_{\ast}$ & $T_2$ & $t_2$ & $N$ & $L$ & $N_t$ & $N_H^{\rm (ini)}$ & $N_H^{\rm (fin)}$  \\ \hline \hline
      1 & 0.1 & 2 & 0.300 & 19.3 & $2048^2$ & 82.62 & 9826 & $49.5^2$ & $2.00^2$ \\ \hline
      2 & 0.5 & 2 & 1.50 & 3.85  & $1024^2$ & 58.42 & 4827 & $35.0^2$ & $2.00^2$ \\ \hline
      3 & 0.5 & 0.02 & 0.875 & 6.60 & $1024^2$ & 58.42 & 4827 & $35.0^2$ & $2.00^2$ \\ \hline
\end{tabular}
\caption{Parameters used in the numerical simulations.  
$m>m_\ast$ and  $m<m_\ast$ correspond to the strongly and weakly interacting regimes, respectively, where we defined $m_\ast$ at Eq.~\eqref{eq:two_vacua}. %
$T_2$ is the temperature of the second phase transition, $t_2$ is the time when the second phase transition occurs, $N$ is the number of grid points, $L$ is the length of the square domain, $N_t$ is the total number of the time steps. $N_H^{\rm (ini)}$ and $N_H^{\rm (fin)}$ are the numbers of horizons included in the simulation domain at the initial time and at the final time, respectively. All the dimensional quantities are normalized by $v_1$.
}
\label{tab:parameters}
\end{table*}

\subsection{Time evolution}

\begin{figure*}[htbp]
    \centering
 \begin{tabular}{ccc}
    \begin{minipage}[]{0.31 \linewidth}
    \centering
        \includegraphics[height=5.4cm, keepaspectratio]{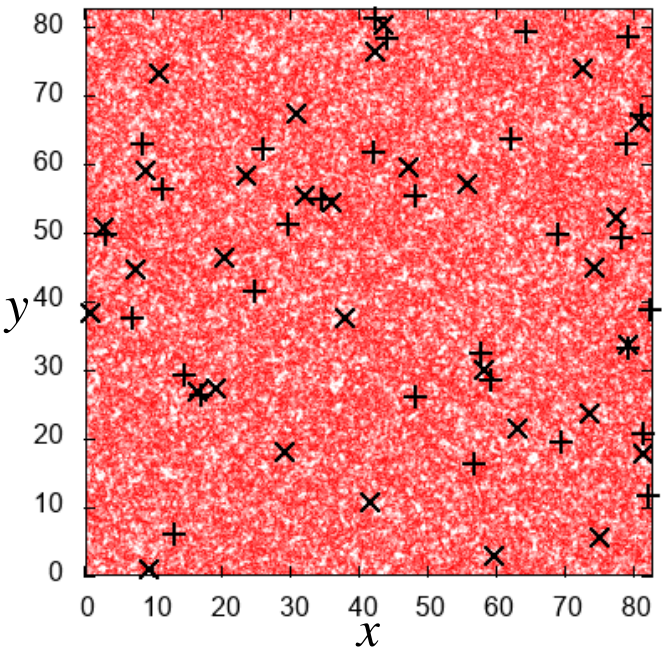}
    \subcaption{$t =  17.51$}
    \label{fig:TE_strong_a}
    \end{minipage}
    &
    \begin{minipage}[]{0.31 \linewidth}
    \centering
    \includegraphics[height=5.4cm, keepaspectratio]{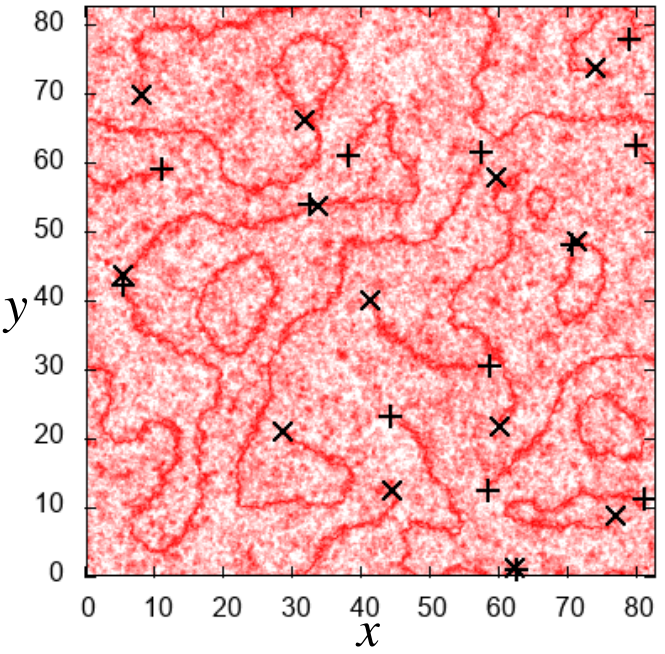}
    \subcaption{$t =31.38$}
    \label{fig:TE_strong_b}
    \end{minipage}
    &
    \begin{minipage}[]{0.35 \linewidth}
    \centering
        \includegraphics[height=5.37cm, keepaspectratio]{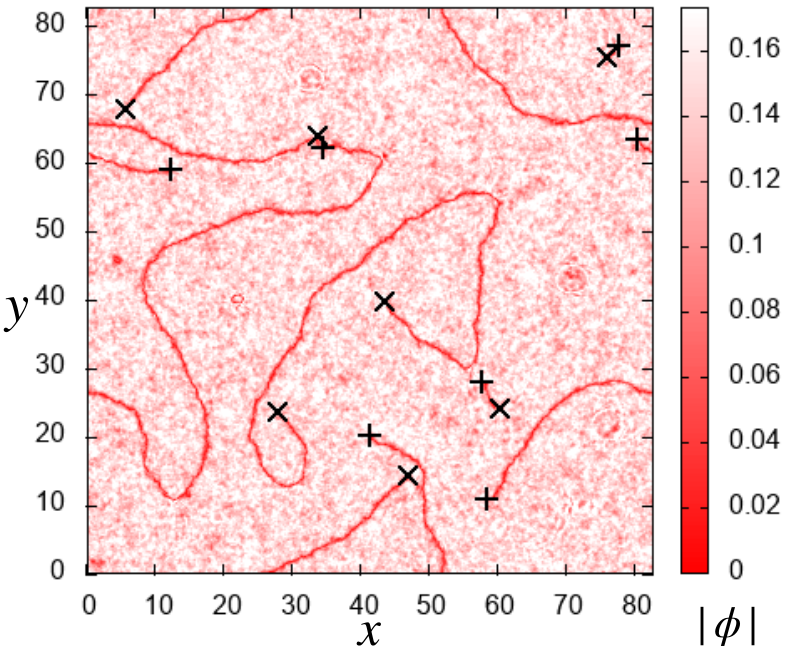}
    \subcaption{$t =41.28$}
    \label{fig:TE_strong_c}
    \end{minipage}
    \end{tabular}
    \caption{[Run~1] Time evolution of the absolute value of $\phi$-field on $x$-$y$ space domain in the strongly interacting regime. The largest value of the color bar is $|\phi|=\tilde{v}_2=0.17 v_1$. Plus/cross signs indicate the centers of 
    $\zeta$-strings with $n_\alpha=1/-1$,
    penetrating $x$-$y$ plane in $z$-direction. Walls are connecting not only pairs of $\zeta$-strings with different $n_\alpha$ but also pairs of $\zeta$-strings with the same $n_\alpha$. All the dimensional quantities are normalized by $v_1$.
}
    \label{fig:TE_strong}
\end{figure*}

In Figs.~\ref{fig:TE_strong_a}-\ref{fig:TE_strong_c}, we show the time evolution of $\phi$
in Run~1, in the range of the strongly interacting regime.
We choose a relatively small~$v_2$, i.e.,  
we assume a relatively large hierarchy between $v_1$ and $v_2$, to clearly distinguish the two phase transitions. 
The plus and cross symbols in Figs.~\ref{fig:TE_strong_a}-\ref{fig:TE_strong_c} indicate the location of 
$\zeta$-strings with $n_\alpha=1$ and $n_\alpha=-1$, respectively.

In Fig.~\ref{fig:TE_strong_a}, we can find 
$\zeta$-strings with $n_\alpha=\pm 1$ are formed after the first phase transition.
On the other hand, 
$\phi$ still fluctuates around $|\phi|=0$ since the second phase transition does not occur yet. Note that $t_2=19.3$ for Run 1.
In Fig.~\ref{fig:TE_strong_b}, the second phase transition has occurred, and walls are formed. We can find that there are $\zeta$-strings at the ends of walls 
and closed walls. %
The former structure is very close to that shown in Fig.~\ref{fig:torus_strong}. 
In Fig.~\ref{fig:TE_strong_c}, we show the field configurations at a late time. We can clearly observe arbitrary combinations of 
$\zeta$-strings with $n_\alpha=\pm 1$ connected by walls, i.e., $w=0\ , \pm 1$ string-walls introduce in Sec.~\ref{sec:string-walls}.
The reconnection of the walls and the merger of the 
$\zeta$-strings with different $n_\alpha$
reduce their numbers as time goes on. For instance, string-wall bends around $(x,y)=(60,60)$ were two string-walls, which are connected by the merger of $\zeta$-strings with different $n_\alpha$. A string-wall around $(x,y)=(60,25)$ was cut off from the longer string-wall extending between $(x,y)=(60,25)$ and $(x,y)=(60,60)$ by the reconnection of the string-walls, corresponding to Fig.~\ref{fig:weak_rec1}. Later, we will explain the reconnection process in the weakly interacting regime in detail.

\begin{figure*}[htbp]
    \centering
 \begin{tabular}{ccc}
    \begin{minipage}[]{0.31 \linewidth}
    \centering
    \includegraphics[height = 5.4cm, keepaspectratio]{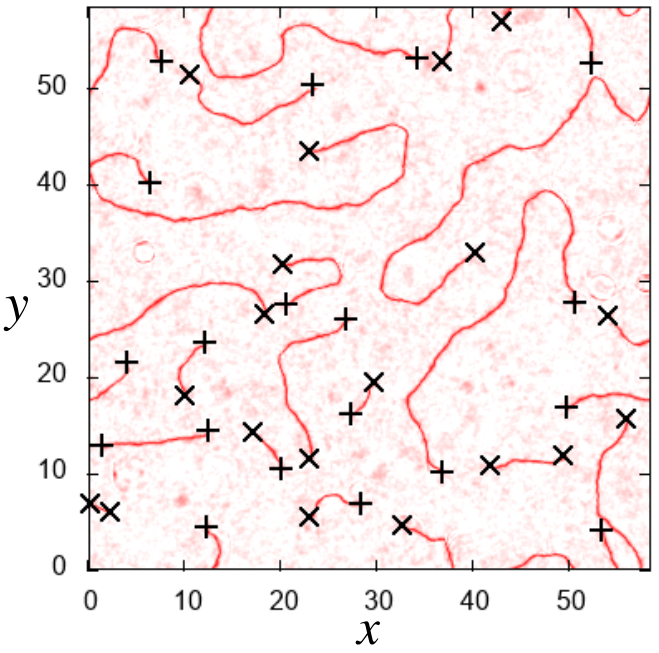}
    \subcaption{$\eta =  14.04$}
    \label{fig:TE_strong_later_a}
    \end{minipage}
    &\begin{minipage}[]{0.31 \linewidth}
    \centering
    \includegraphics[height = 5.4cm, keepaspectratio]{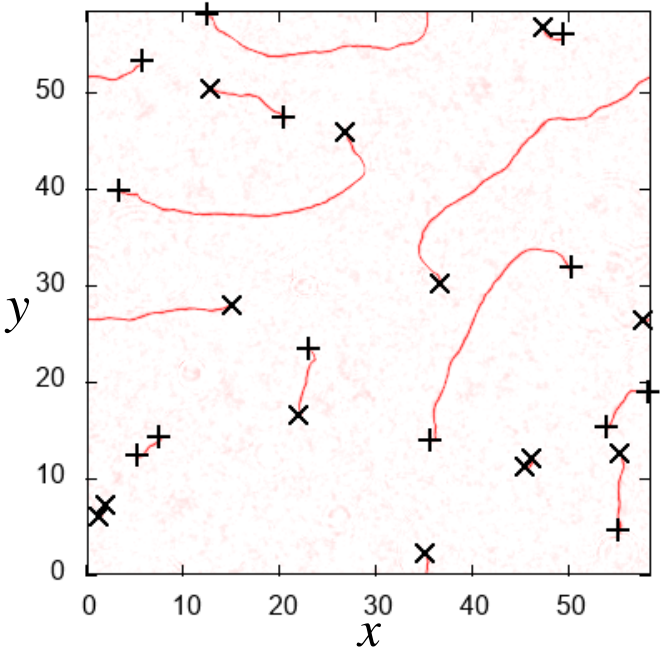}
    \subcaption{$\eta =  20.92$}
    \label{fig:TE_strong_later_b}
    \end{minipage}
    &
    \begin{minipage}[]{0.35 \linewidth}
    \centering
    \includegraphics[height = 5.37cm, keepaspectratio]{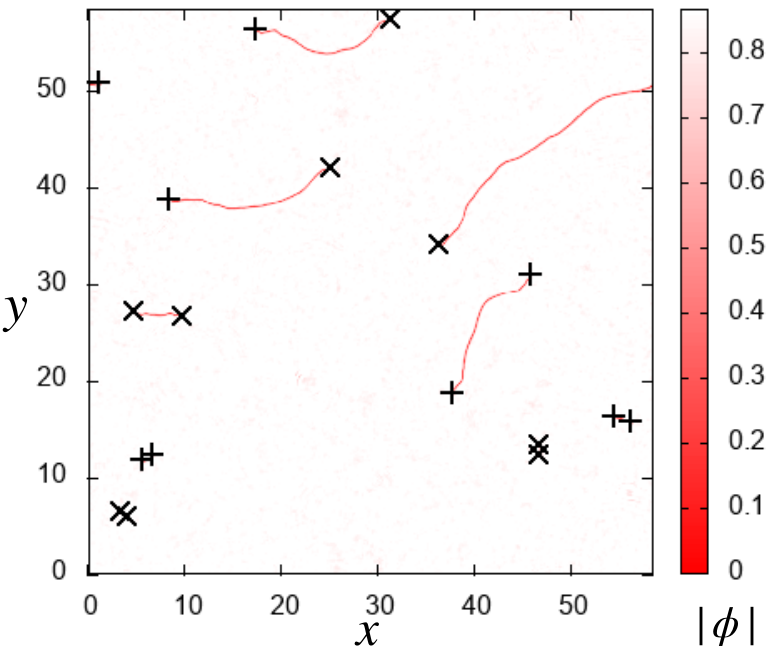}
    \subcaption{$\eta =  27.79$}
    \label{fig:TE_strong_later_c}
    \end{minipage}
    \end{tabular}
    \caption{[Run 2] Time evolution of the absolute value of $\phi$-field in strongly interacting regime after the formation of walls. See the caption of Fig.~\ref{fig:TE_strong}. The largest value of the color bar is $|\phi|=\tilde{v}_2=0.87 v_1$. Walls pull $\zeta$-strings to closer and stretch straight. $w=0$ string-walls vanish, while a $w=\pm 1$ string-walls do not vanish and twirl each other. All the dimensional quantities are normalized by $v_1$.}
    \label{fig:TE_strong_later}
\end{figure*}

In Figs.~\ref{fig:TE_strong_later_a}-\ref{fig:TE_strong_later_c}, we show the case of Run~2, again in the range of the strongly interacting regime. To see the time evolution of the string-wall system for a longer time after the wall formation, we choose a relatively large $v_2$, i.e., we assume a relatively small hierarchy between $v_1$ and $v_2$. All panels are the field configurations after the second phase transition ($t_2=3.85$). 

Long after the second phase transition, the numbers of $\zeta$-strings and the walls are further reduced and the walls are more straightened by their tension. 
One of the reasons for the decrease in the number is that
$w=0$ string-walls, walls with two  $\zeta$-strings with different $n_\alpha$ attached, vanish after their shrinking, as we can see, e.g., around $(x,y)=(23,20)$ in Figs.~\ref{fig:TE_strong_later_b} and \ref{fig:TE_strong_later_c}. 
In contrast, $w=\pm 1$ string-walls, walls with two $\zeta$-strings with the same $n_\alpha$, survive after their shrinking and the $\zeta$-strings revolve around each other, as we see it around $(x,y)=(2,7)$ in Figs.~\ref{fig:TE_strong_later_a}-\ref{fig:TE_strong_later_c}. That is because the $w=\pm 1$ string-walls are topologically stable, as discussed in Sec.~\ref{sec:string-walls} and shown numerically in Figs.~\ref{fig:torus_strong} and \ref{fig:torus_strong2}.

\begin{figure*}[htbp]
    \centering
 \begin{tabular}{ccc}
    \\
    \begin{minipage}[]{0.31 \linewidth}
    \includegraphics[height = 5.4cm, keepaspectratio]{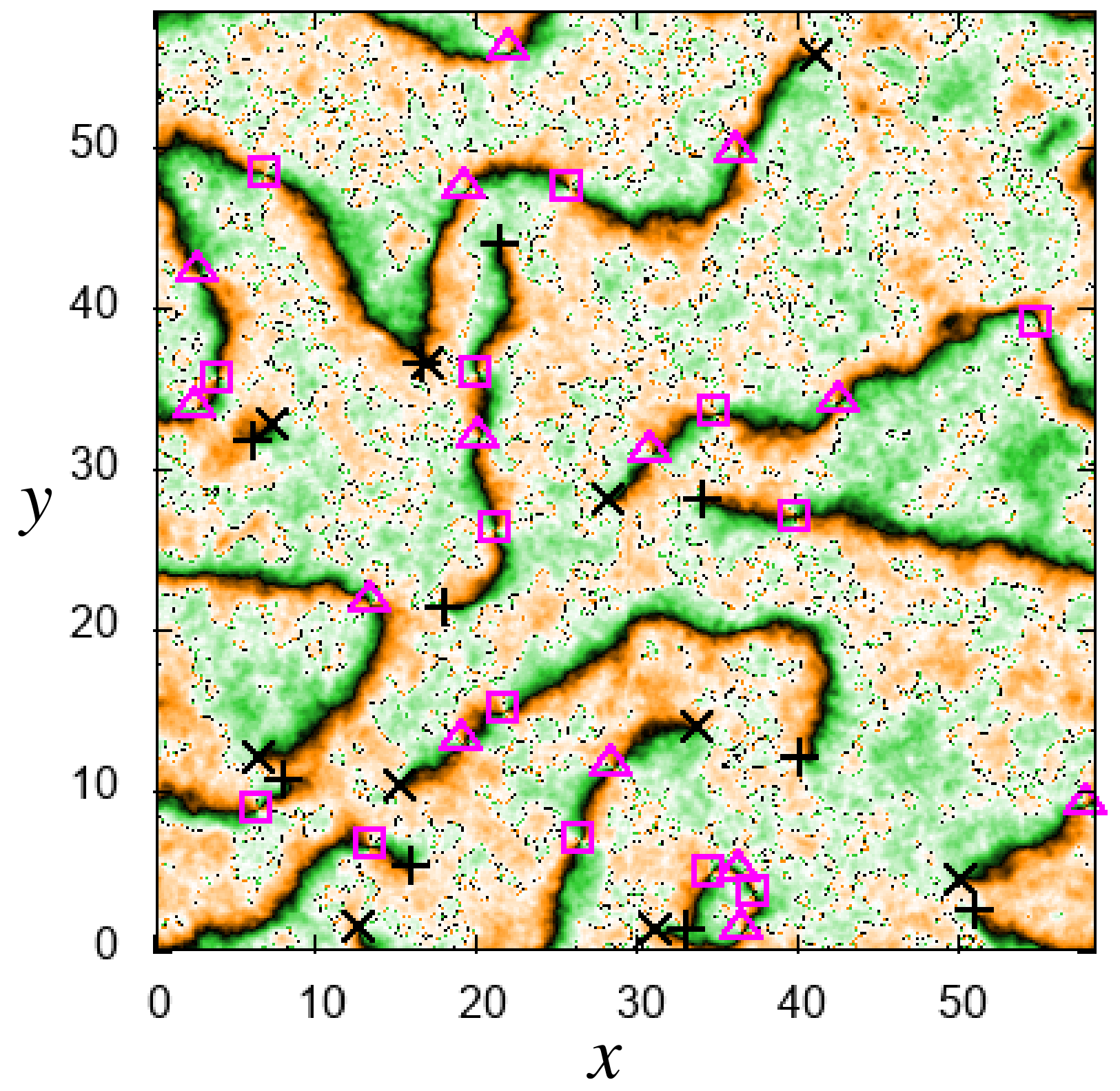}
    \subcaption{$\eta =  19.54$}
    \label{fig:TE_weak_d}
    \end{minipage}
    &\begin{minipage}[]{0.31 \linewidth}
    \includegraphics[height = 5.4cm, keepaspectratio]{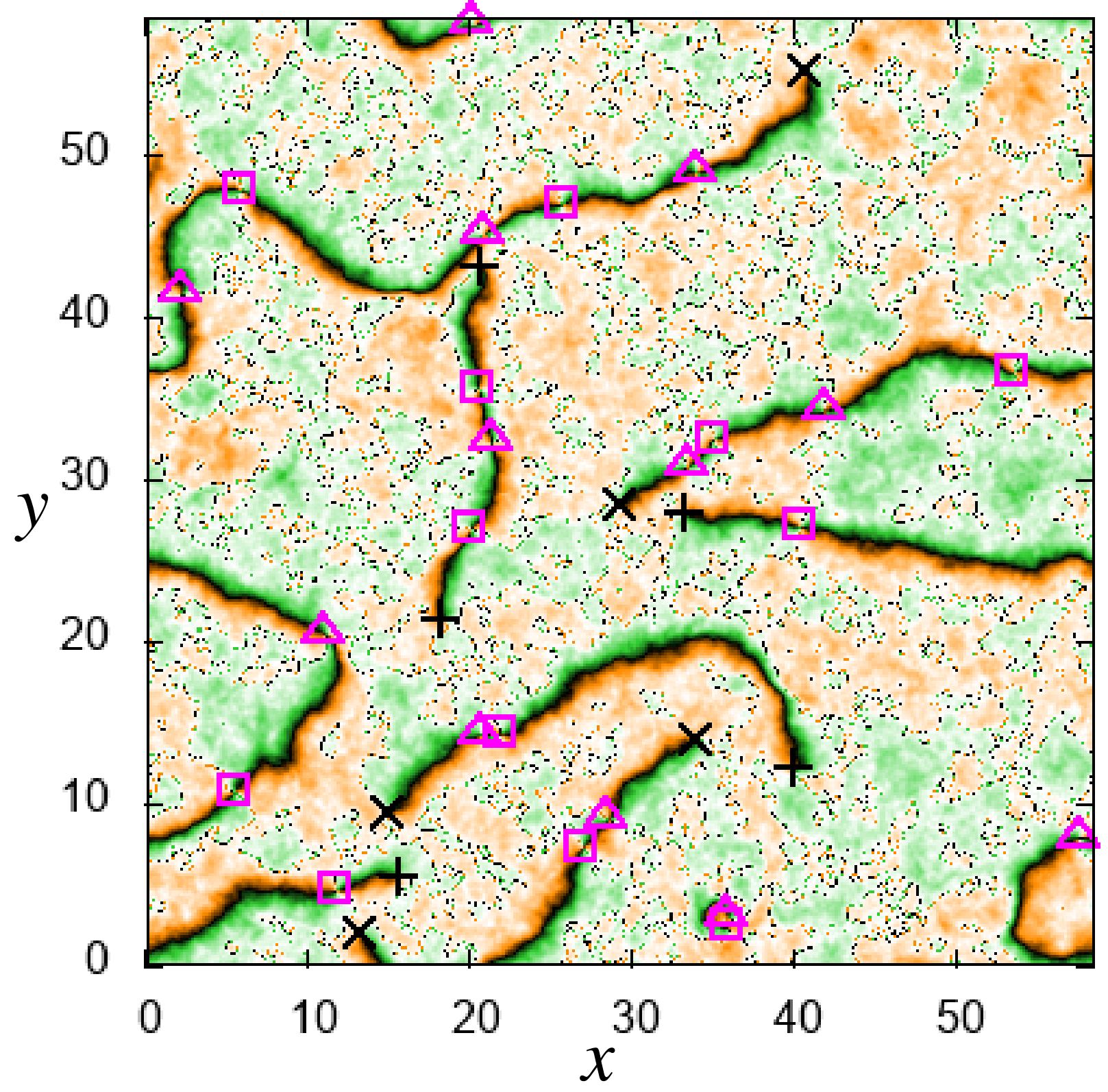}
    \subcaption{$\eta =  25.04$}
    \label{fig:TE_weak_e}
    \end{minipage}
    &
    \begin{minipage}[]{0.35 \linewidth}
    \includegraphics[height = 5.37cm, keepaspectratio]{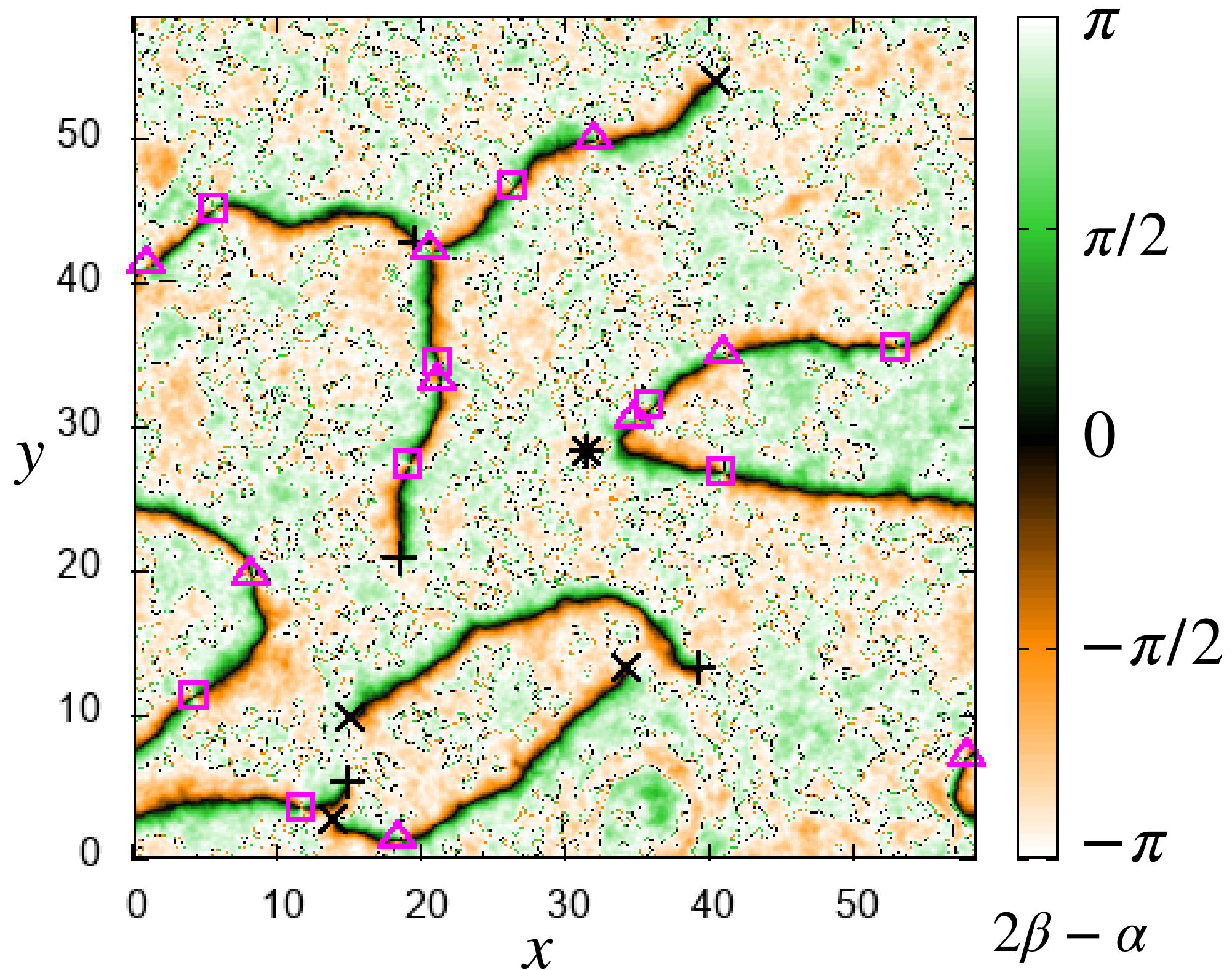}
    \subcaption{$\eta =  29.17$}
    \label{fig:TE_weak_f}
    \end{minipage}
    \end{tabular}
    \caption{ %
    [Run 3] Time evolution of the relative argument, $2\beta-\alpha$, 
    in the weakly interacting regime. See Fig.~\ref{fig:weak_phase} for the meaning of the colors. The black color shows the center of the walls. The white color (including noisy dots) corresponds to the vacuum. Plus/Cross signs indicate centers of $\zeta$-strings with $n_\alpha=1/-1$, penetrating $x$-$y$ plane along $z$-axis.  Rectangle/triangle represents $\phi$-string with $\tilde{n}_\beta=-1/1$.
    The width of the $\zeta$-strings is smaller than these plus/cross symbols. Walls are connecting not only pairs of $\zeta$-strings with different $n_\alpha$ but also pairs of $\zeta$-strings with the same $n_\alpha$. The width of walls is much larger than that of $\phi$-strings as evaluated at sec.~\ref{sec:string-walls}, and thus, the walls look pinched at the location where $\phi$-strings exist. Walls are more diffuse than those in the strongly interacting regime. The small noisy dots between strings are numerical artifacts coming from the discontinuity of the arguments of the fields. All the dimensional quantities are normalized by $v_1$.}
    \label{fig:TE_weak}
\end{figure*}

In Figs.~\ref{fig:TE_weak_d}-\ref{fig:TE_weak_f}, we show the case of Run~3, corresponding to the weakly interacting regime. 
In this regime,
the angle, $2\beta-\alpha$, is more useful to specify a wall than the absolute value, $|\phi|$, as we explained in Sec.~\ref{sec:walls}. 
The black  region indicates $2\beta-\alpha = 0 \mod 2\pi$, corresponding to the bumps on the torus Eq.~\eqref{eq:bump}, i.e., the center of the walls.

We eliminate the earlier frames since,
qualitatively, the behavior of the string-wall system looks the same as in the strongly interacting regime.
All the frames shown in Fig.~\ref{fig:TE_weak} are after the second phase transition.
There are $w=0$ string-walls and $w=\pm 1$ string-walls,
again consistent with the explanation in Sec.~\ref{sec:string-walls}.
In contrast to Run~1 and Run~2, the walls are more diffuse and their tension is weaker. As a result, their straightening and shrinking are much slower. 
Two disconnected string-walls with the ends of $\zeta$-strings around $(x,y)=(50,5)$ in Fig.~\ref{fig:TE_weak_d} become one connected string-wall in Fig.~\ref{fig:TE_weak_e} by the merger of  $\zeta$-strings with different $n_\alpha$. The same can also be seen around 
$(x,y)=(5,10)$ 
in Fig.~\ref{fig:TE_weak_d},
which merges in Fig.~\ref{fig:TE_weak_e}.
Similarly, $\phi$-strings with different $\tilde{n}_\beta$ merge with each other on walls, which are observed around $(x,y)=(5,35)$ from Fig.~\ref{fig:TE_weak_d} to Fig.~\ref{fig:TE_weak_e},
and around $(x,y)=(20,15)$ and $(28,8)$ from Fig.~\ref{fig:TE_weak_e} to Fig.~\ref{fig:TE_weak_f}.
We also find ripples in the bulk, which are remnants of the annihilation of 
$\zeta$-strings with different $n_\alpha$, seen, e.g., around $(x,y)=(5,10)$, $(5,30)$ and $(50,5)$ between Fig.~\ref{fig:TE_weak_d} and Fig.~\ref{fig:TE_weak_e}.
Collisions of $\phi$-strings with different $\tilde{n}_\beta$ also result in the generation of ripples as seen, e.g., around $(x,y)=(35,5)$, $(28,8)$ and $(20,15)$ between Fig.~\ref{fig:TE_weak_e} and Fig.~\ref{fig:TE_weak_f}.

\begin{figure}
    \centering
        \vspace*{-5mm}
    \centering
        \includegraphics[height = 3cm, keepaspectratio]{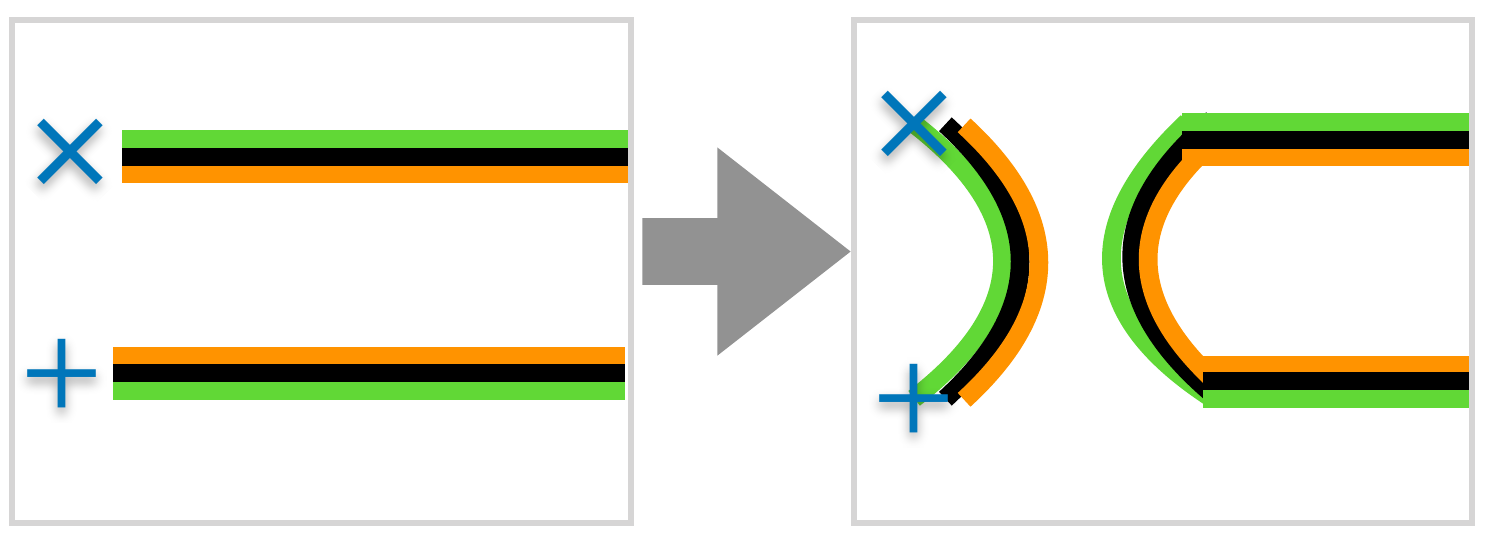}
    \caption{Schematic diagram to explain the reconnection happening around $(x,y)=(35,30)$ in Fig.~\ref{fig:TE_weak}. From Fig.~\ref{fig:TE_weak_e} to Fig.~\ref{fig:TE_weak_f}, two string-walls are approaching each other, and the walls are reconnected. The $\zeta$-strings with different $n_\alpha$ are connected by a short wall, approach each other and ought to be annihilated. On the right-hand side, the rest parts of the string-walls are connected and become a longer string-wall.
    The blue plus/minus signs represent $\zeta$-strings with $n_\alpha=1/-1$. 
    }
    \label{fig:weak_rec2}
\end{figure}

In addition, we observe reconnection around $(x,y)=(20,40)$ from Fig.~\ref{fig:TE_weak_e} to Fig.~\ref{fig:TE_weak_f}.~\footnote{It somewhat looks a three-way branch at the place, but it is impossible to form one in our double charged model. In the triple charged model, where we have two scalar fields and one of these has a global charge three times as large as the other, three-way branches appear.} See Fig.~\ref{fig:s1_slow} for the snapshots between these frames and Fig.~\ref{fig:weak_rec1} for the schematic diagram. 
Another reconnection happens around $(x,y)=(35,30)$ from Fig.~\ref{fig:TE_weak_e} to Fig.~\ref{fig:TE_weak_f}. See Fig.~\ref{fig:s2_slow} for the snapshots between these frames and Fig.~\ref{fig:weak_rec2} for the schematic diagram. According to the frames between Fig.~\ref{fig:TE_weak_e} and Fig.~\ref{fig:TE_weak_f}, it seems that this reconnection is triggered by the attractive force between the $\zeta$-strings with different $n_\alpha$, and the walls themselves interact and reconnect with each other. It should be noted that the former type of reconnection changes not the string-wall number but the distribution of the wall length, while the latter type reduces the string-wall number by one and forms a longer wall. Reconnection between $w=1$ string-wall and $w=-1$ string-wall reduces the total number of $w=\pm 1$ string-walls in equilibrium (between the wall tension and the repulsive force between $\zeta$-strings) at the late time, while that between $w=0$ string-walls increases the total number. If either type of reconnection happens for the same string-wall, it can form a loop, (though the former type possibly does not form a loop.)
Therefore, the reconnection processes can advance the loop formation and decay of the defects. These are more complicated in three-dimensional space and three-dimensional simulation is required to determine if the distributions of walls and strings approach the scaling law.

\begin{figure}
    \centering
        \vspace*{-5mm}
    \centering
        \includegraphics[height = 6.4cm, keepaspectratio]{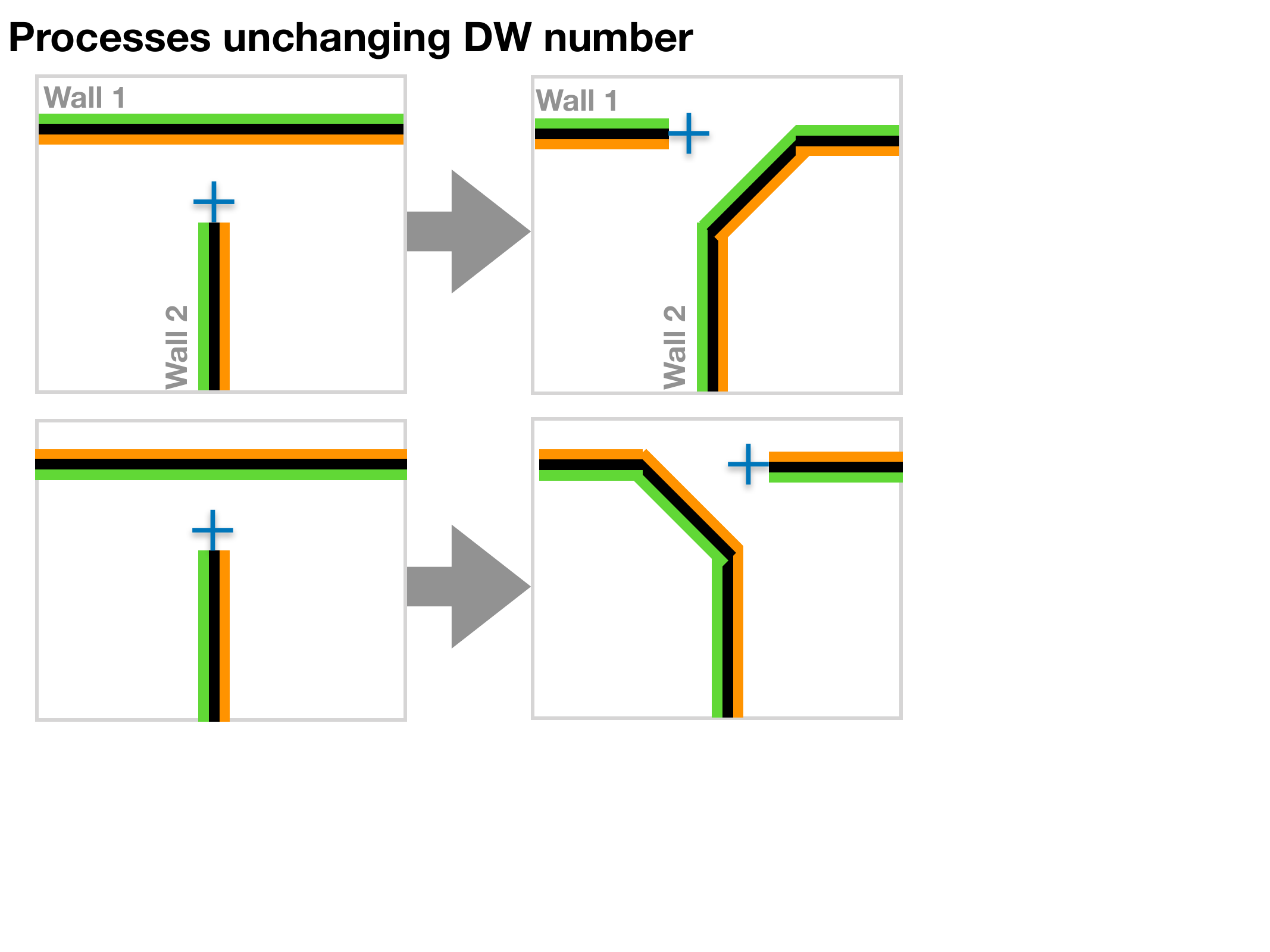}
    \caption{The top panel is a schematic diagram to explain the reconnection happening around $(x,y)=(20,40)$ in Fig.~\ref{fig:TE_weak}.  %
    In the left panel,
    two walls are approaching each other and one wall is ended by a $\zeta$-string with $n_\alpha=1$ represented by '$+$'.
    Then, Wall 1 is broken in the middle and the string is rejoined to the left end, and the right rest part of Wall 1 is rejoined to Wall 2 as shown in the right panel.
    The bottom panel shows another possible reconnection if Wall 2 approaches Wall 1 from the opposite direction (The color pattern of Wall 1 is upside down).} 
    \label{fig:weak_rec1}
\end{figure}

\subsection{Structure of string-wall system}

\begin{figure*}
    \centering
        \vspace*{-5mm}
    \centering
        \includegraphics[height = 10cm, keepaspectratio]{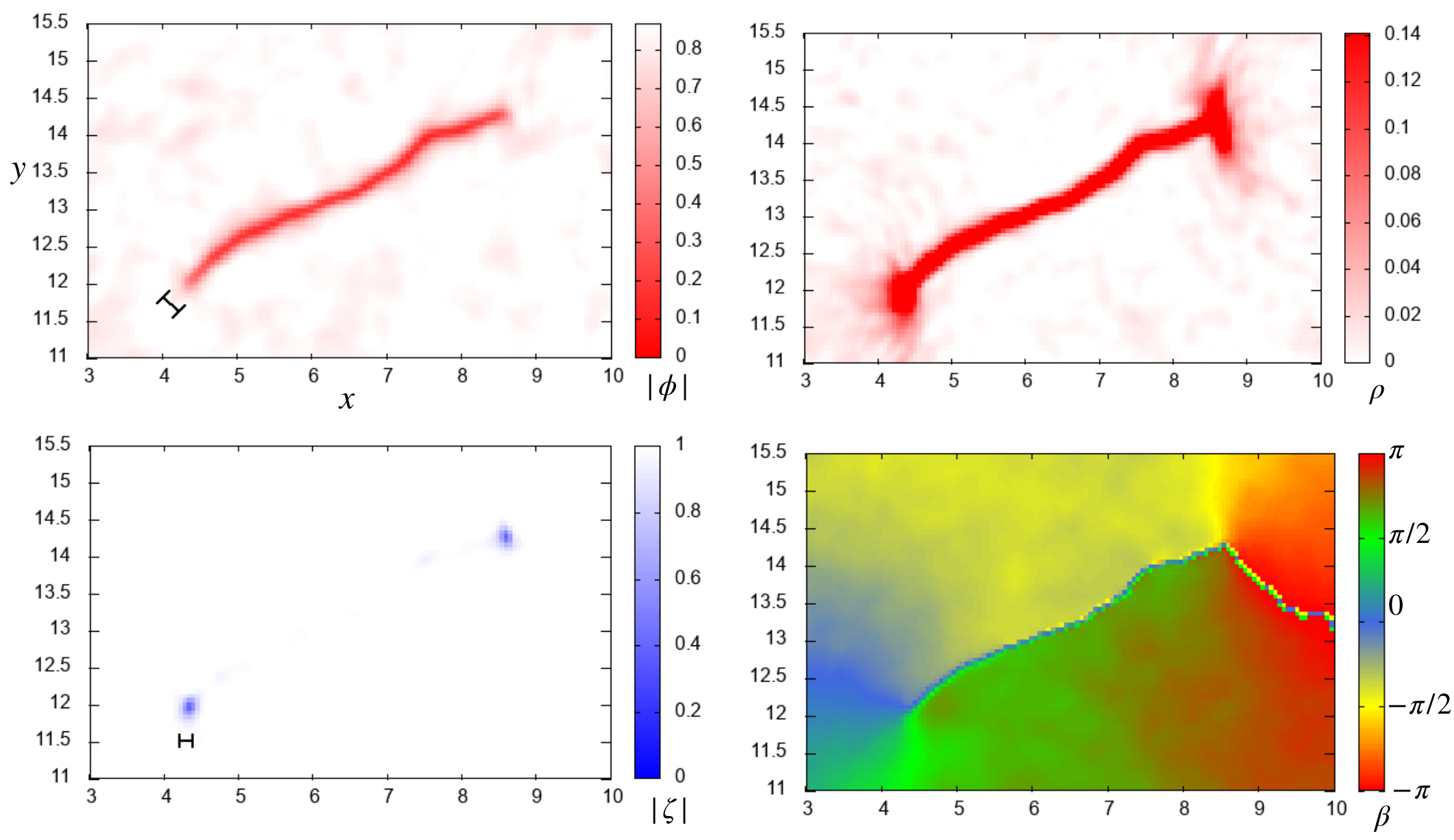}
    \caption{[Run 2] A $w=1$ string-wall located around $(x,y)=(7,13)$ at $t=19.54 v_1^{-1}$. (Top-left) Absolute value of $\phi$-field, $|\phi|$. The red area corresponds to the center of walls and the white area corresponds to the vacuum. The largest value of the color bar is $|\phi|=\tilde{v}_2=0.87 v_1$. 
    The indicated length is twice of the wall-width length scale~$\delta_{\rm DW}^{\rm (strong)}/a =0.14 v_1^{-1}$. (Top-right) Energy density. The maximum value of the color bar is the relative energy density at the torus core to the vacuum energy density, $0.14 v_1^4$.  (Bottom-left) Absolute value of $\zeta$-field, $|\zeta|$. The maximum value of the color bar is $|\zeta|=v_1$. The indicated length is twice of the $\zeta$-string width length scale~$\delta_{\zeta}/a=0.085 v_1^{-1}$. The width of  $\zeta$-strings located at the ends of walls is comparable with the wall width. (Bottom-right) Argument of $\phi$-field. The $\phi$-argument continuously advances by $\pi$ around an end of a wall. The $\phi$-argument almost discontinuously changes by $\pi$ when crossing a wall. 
    The small noisy dots emanating from one of the $\zeta$-strings are numerical artifacts coming from the discontinuity of the arguments of the fields. All the dimensional quantities are normalized by $v_1$ in these panels. }%
    \label{fig:STR_strong}
\end{figure*}

In this subsection, we see the structure of the string-walls more in detail. %
We focus on a $w=1$ string-wall found in Run 2 in the strongly interacting regime and show the energy density, the absolute value of $\phi$, that of $\zeta$, and the argument of $\phi$, $\beta$, in Fig.~\ref{fig:STR_strong}.
The corresponding time evolution has been shown in Figs.~\ref{fig:TE_strong_later_a}-\ref{fig:TE_strong_later_c}.

\begin{figure*}
    \centering
        \vspace*{-5mm}
    \centering
        \includegraphics[height = 14cm, keepaspectratio]{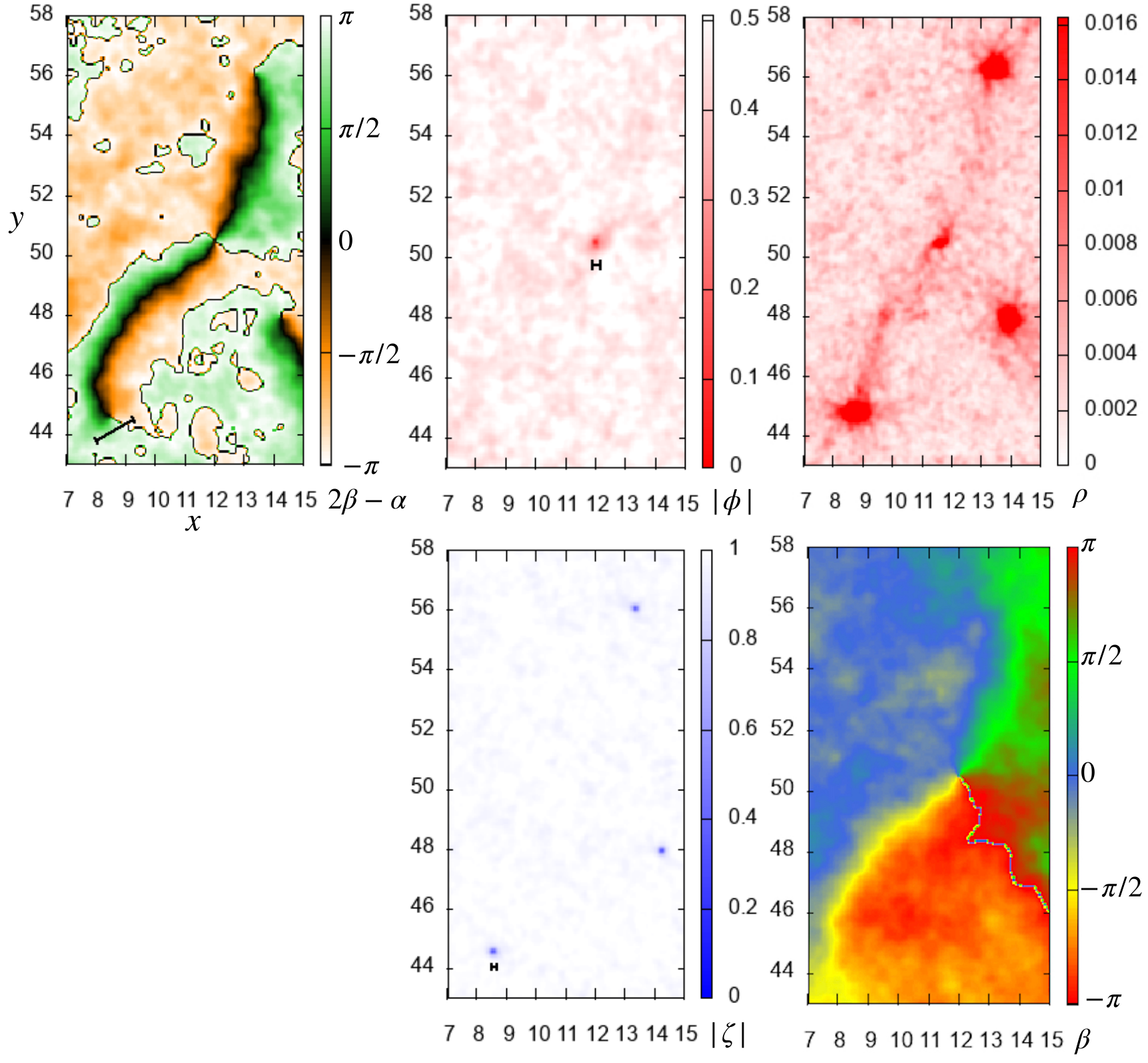}
    \caption{ [Run 3] A $w=-1$ string-wall located around $(x,y)=(11,50)$ at $t = 23.67 v_1^{-1}$. (Top-left) Twice of the $\phi$-argument subtracted by the $\zeta$-argument, $2\beta-\alpha$. The black area corresponds to the center of walls, i.e., $0 \mod 2\pi$ %
    and the white area corresponds to the vacuum, i.e., $\pi \mod 2\pi$. %
    The indicated length is twice as large as the wall width, $\delta_{\rm DW}^{\rm (weak)}/a= 0.70 v_1^{-1}$. The color change from green to orange through black indicates the direction on the torus in $XYZ$-space. (Top-middle) The absolute value of $\phi$-field. The white region corresponds to the field value at the vacuum, $|\phi|=\tilde{v}_2=0.50 v_1$, and the red region corresponds to $|\phi|=0$. The large red dot corresponds to a $\phi$-string with $n_\beta=-1$. 
    The indicated length is twice of the $\phi$-string width length scale, $\delta_{\phi}^{\rm (weak)}/a =0.14 v_1^{-1}$.
    (Top-right) Energy density. The maximum value is the relative energy density at the torus core to the vacuum energy density, 
    $0.016 v_1^4$. (Bottom-middle) Absolute value of $\zeta$-field. The white region corresponds to the vacuum, $|\zeta|= v_1$, and the blue region corresponds to $|\zeta|=0$. The blue large dots around $(x,y)=(8,44)$ and $(13,56)$ corresponds to $\zeta$-strings with $n_\alpha=-1$. 
    The indicated length is twice of the $\zeta$-string width length scale, $\delta_{\zeta}/a =0.070 v_1^{-1}$. (Bottom-right) Argument of $\phi$-field. The $\phi$-argument continuously advances by $2 \pi$ around a $\phi$-string. The $\phi$-argument continuously changes by $\pi$ when crossing a wall. %
    In the top-left and the bottom-right panels, the small noisy dots are numerical artifacts coming from the discontinuity of the arguments of the fields. All the dimensional quantities are normalized by $v_1$.}
    \label{fig:STR_weak}
\end{figure*}

The wall is sharper compared to that in Fig.~\ref{fig:STR_weak} because the energy difference between the vacuum and the torus core %
is higher than that between the vacuum and the bump on the torus for the weakly interacting regime. (See Fig.~\ref{fig:pot_torus_2_strong}.) 
The top-left panel shows the absolute value of $\phi$-field, $|\phi|$, as in Figs.~\ref{fig:TE_strong} and \ref{fig:TE_strong_later}.  
The top-right panel shows the relative energy density to the vacuum energy density.
The bottom-left panel shows the absolute value of $\zeta$-field, $|\zeta|$. By comparing  these two panels, we see the wall width and the core width of $\zeta$-strings at the end of the walls are comparable to each other as we saw in Eq.~\eqref{eq:strong_width_comp} in Sec.~\ref{sec:walls}. 
The bottom-right panel shows the argument of $\phi$-field, $\beta$. We see that $\beta$ quickly advances by $\pi$ when crossing the wall because $\phi$-field passes near the torus core (the origin on $\alpha$-constant surface) in $XYZ$-space, predicted in Sec.~\ref{sec:walls}.
Furthermore, $\phi$-argument advances by $\pi$ when going around a $\zeta$-string as shown in Table~\ref{tab:chages} since \eqref{eq:vac_rel} holds in a vacuum.

Fig.~\ref{fig:STR_weak} shows snapshots of a $w=-1$ string-wall found in Run~3 in the weakly interacting regime. 
The top-right panel shows the relative energy density. The wall is less sharp and has less energy compared to Fig.~\ref{fig:STR_strong} because the two vacua on $\alpha$-constant surface are separated only by low local maxima away from the torus core in $XYZ$-space. (Fig.~\ref{fig:pot_torus_2_weak}.) 
The top-middle panel shows the absolute value of $\phi$-field.
We can see a $\phi$-string, where $\phi$-field surrounds the torus core in $XYZ$-space. As expected in Sec.~\ref{sec:walls}, it is hard to see the wall in this plot in contrast to the case of the strongly interacting regime. 
The top-left panel shows the relative argument, $2\beta-\alpha$. The center of the walls is located at $2\beta-\alpha=0 \mod 2\pi$, while the vacuum is located at $2\beta-\alpha = \pi \mod 2\pi$. 
We see that the string-wall includes $\phi$-string on the wall between the two $\zeta$-strings.
The wall width is much larger than the width of $\phi$-strings in field-value-based view as we saw in Eq.~\eqref{eq:dw_width_weak_limit} in Sec.~\ref{sec:string-walls}, and thus, the wall look pinched at the location of $\phi$-string. 
The bottom-right panel shows the argument of $\phi$-field, $\beta$. 
We see the phase gradually advances by $\pi$ when crossing the wall because $\phi$-field does not need to pass the torus core but can pass the surface of the torus away from the torus core along $\beta$-cycle in $XYZ$-space (See Fig.~\ref{fig:pot_torus_2_weak}.) Furthermore, $\beta$ advances by $-2 \pi$ when going around the  $\phi$-string.

\section{Summary}

The condensation of integer multiple charged fields under the same continuous symmetry easily leads to a discrete symmetry. If the fields condensate in order, it can lead to a sequence of the symmetry-breakings with intermediate discrete symmetries. Such examples are found in $SO(10)$ grand unified theories and in the promotion of $U(1)_{B-L}$ symmetry. One of the features of the symmetry-breaking chains is to give a type of hybrid defect, 
string-wall configuration, which is possibly a source of the primordial gravitational waves.

As a simplest model, we incorporated two scalar fields to form one global $U(1)$ symmetry where one scalar field has the double charge of the other one. By assuming the condensation energy scale of the double-charged field is higher than that of the single-charged field, %
 $U(1)$ breaks into the trivial group through the intermediate symmetry~$Z_2$, leading to a string-wall system. The importance of the gauge field will be discussed in another publication.

By analytical consideration, we clarified the vacuum structure of this model in low energies is a torus knot winding around a torus. 
We found that 
the string-wall configurations are very different in two distinct regimes, the strongly interacting regime and the weakly interacting regime. 
These two regimes are distinguished by %
the strength of the interaction between the two fields, $\zeta$-field (double-charged field) and $\phi$-field (single-charged field).
In the strongly interacting regime, $\phi$-strings do not exist, and either a pair of $\zeta$-strings with different $n_\alpha$ or those with the same $n_\alpha$ are connected by a wall.
In the weakly interacting regime, $\phi$-strings do exist, and a pair of  $\zeta$-strings with the same $n_\alpha$ need to be connected to a $\phi$-string by walls, though a pair of $\zeta$-strings with different $n_\alpha$ can be connected by a wall.
It is notable that the structure of wall solutions is very different between the two regimes.
We found the wall solutions in both regimes and explained the walls are thinner and sharper, and have more energy in the strongly interacting regime. 
These differences in the existence of $\phi$-strings and in the structure of walls originate in the difference in the potential structures. The potential in the strongly interacting regime does not admit $\phi$-string because of the absence of $S^1$ structure along $\arg \phi$-direction. 
The largeness of the energy difference between the vacuum and the torus core leads to walls with high-energy and the passage near the torus core leads to an almost discontinuous change of the argument of $\phi$-field, resulting in the sharpness of the walls.
In the weakly interacting regime, since $S^1$ structure along $\arg \phi$-direction approximately remains, 
$\phi$-strings exist. The bumps on the $S^1$ yields two walls around a $\phi$-string and the farness of these bumps on the torus surface away from the torus core results in the gradual change of $\arg \phi$ and contributes to the faintness of the walls.
We also compared the width of $\zeta$-strings, $\phi$-strings and walls in both regimes. 

We confirmed our %
analytical predictions
through two-dimensional cosmological simulations. We observed that the $\zeta$-strings with different $n_\alpha$ are connected by walls and are finally annihilated. In addition, we observed that the $\zeta$-strings with the same $n_\alpha$ are connected by walls and form a bound state with $w=\pm 1$. We also found a novel reconnection of string-walls in our model.

\section{Discussion}

If $\zeta$-field is not a field but a constant, our model is reduced to the axion model with $N=2$, whereas if $\phi$-field is not a field but a constant, our model is reduced to the axion model with $N=1$. Therefore, our model can be considered as an extension of the axion model. %
It is remarkable that our model is expected to be safe from the domain-wall problem unlike usual $N\geq 2$ axionic models. The reason is that each wall is ended by $\zeta$-strings except for closed walls in our model. Suppose that we consider $\phi$-field as the axion. $\phi$-string is connected to two walls (as in $N=2$ axion model) but each wall is ended by a $\zeta$-string in the weakly interacting regime. Also in the strongly interacting regime, a wall is ended by two $\zeta$-strings. Since the domain-wall problem for $N\geq 2$ axion models originates in the fact that the walls do not have ``the ends'', it is natural to expect that our model is free from the domain-wall problem.
Therefore, our model suggests that the promotion of the coefficient of the axion potential term to a field 
can be a possible way to avoid the domain-wall problem found in the $N=2$ axion model.
Our model can be generalized to the cases with $N\geq 3$ by hiring $\zeta$-field with the charge of an arbitrary integer multiple of that of $\phi$-field, leading to the sequence of $U(1)\rightarrow Z_N \rightarrow I$.

\begin{figure}
    \centering
        \vspace*{-5mm}
    \centering
        \includegraphics[height = 8cm, keepaspectratio]{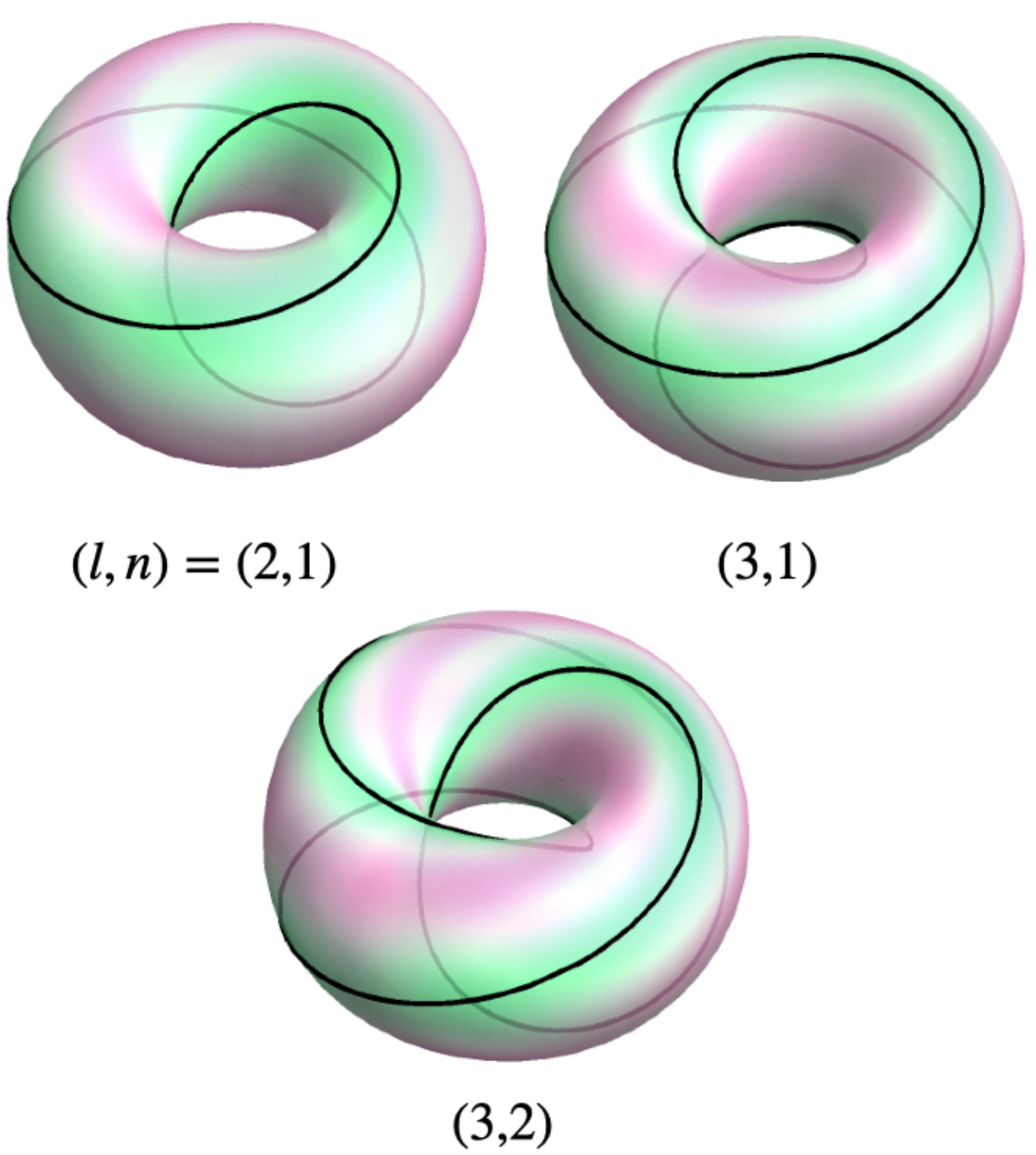}
    \caption{Vacuum structure for $(l,n)=(2,1), (3,1)$ and $(3,2)$. The radii and the angles are explained at Fig.~\ref{fig:XYZ}. The color on the surface represents the potential energy, where pink color means higher potential energy than green color. The black line corresponds to the vacuum manifold in each model. $(l,n)=(2,1)$ (top-left panel) is the model that we have been discussing in this paper and the vacuum manifold is TN(2,1). Similarly,  the vacuum manifold for $(l,n)=(3,1)$ model (top-right panel) and $(l,n)=(3,2)$ model (bottom panel) are  TN(3,1) and TN(3,2), respectively. The latter is called a trefoil knot.}
        \label{fig:torus_31_32}
\end{figure}

It is known that some solitons have knot structures in the physical space. It is remarkable, however, the solitons found in our simple model with two differently charged fields have the torus-knot structure in the field space (see Fig.~\ref{fig:torus_31_32}). Of course, our soliton mapped into the physical space can be either unknotted or knotted irrespective of the vacuum structure. Including other cases with $(l,n)=(3,1)$ and $(3,2)$, where $l$ is the global charge for $\zeta$-field and $n$ is that for $\phi$-field, the vacuum structure is found to be TN$(l,n)$. We note that the strongly interacting regime exists only approximately %
for $(l,n)=(3,1)$ and $(3,2)$ because the Hessian at the torus core in the field space cannot be negative.

\begin{figure}
    \centering
        \vspace*{-5mm}
    \centering
        \includegraphics[height = 7.5cm, keepaspectratio]{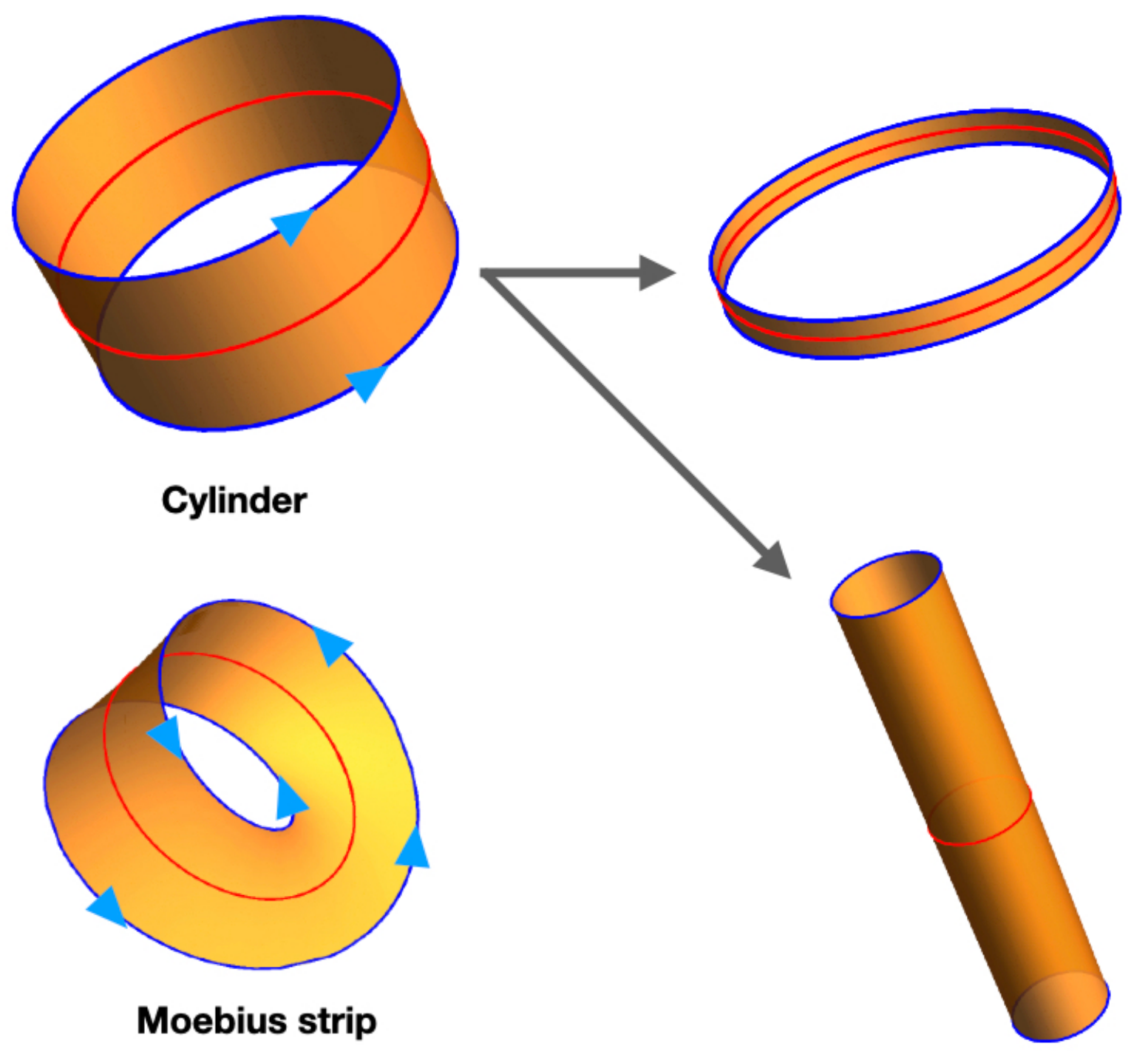}
    \caption{String-wall configurations and the evolution in the three-dimensional physical space. The yellow strips are walls. The blue curves are $\zeta$-strings and the red curves are $\phi$-strings. The blue triangle indicates the directions of the $\alpha$-cycle. }
        \label{fig:development_3d}
\end{figure}

In three-dimensional space, the string-wall configurations seen in our simulations are expected to have richer structures. One string-wall can look a cylinder made of a ``ribbon'' of a wall with two ends of $\zeta$-strings. (See Fig.~\ref{fig:development_3d}.) Another is expected to look a Moebius strip with one end of a $\zeta$-string. In addition, it would be interesting to examine how they evolve in three-dimensional simulations; if a cylinder becomes thinner because of the wall tension or it becomes longer because of the repulsive force between the two $\zeta$-strings. A Moebius strip will involve more complicated processes.

A model with differently charged two scalar fields was discussed in \cite{Hiramatsu:2019tua,Hiramatsu:2020zlp}. In these works, they have a $U(1)_{\rm gauge}\times U(1)_{\rm global}$ symmetry, where $U(1)_{\rm global}$ is an accidental symmetry appearing because they do not have interaction terms like those proportional to $m$ in our model. Unlike our model, with the accidental symmetry, they do not find any wall solutions. 

Another important topic is whether this system has the scaling property. The scaling property of string-wall networks can be an important issue to judge if it does not disagree with the observational necessary condition that both strings and walls should not overdominate the Universe. Since walls in our model are nontopological, $w=0$ string-walls vanish and $w=\pm 1$ string-walls remain as two bound strings which are regarded as effectively $n_\alpha=\pm 2$ strings at the late time, the energy density of the defects seems to behave as normal strings with higher winding numbers.
Our results are the first step toward the three-dimensional simulation for the quantitative estimation of the correlation length of the string-wall network. The gauge field blocks the long-range force between the defects beyond the Compton length. %
Therefore, the simulation including the gauge field in the three-dimensional box is necessary and is left for a future study.

The gravitational wave signature of string-wall systems will give a test of our model. Since the formation of strings with higher windings requires higher energy, most of the population is supposed to be dominated by strings connected by walls, as supported by our simulation. The gravitational wave spectrum from string-wall systems was evaluated in \cite{Dunsky:2021tih} in Nambu-Goto description. This work suggests that if the energy density of the walls is larger than the string energy density at the wall formation, there is a bump-like feature at high frequency region in the gravitational wave spectrum. This situation occurs if the energy scales of the first phase transition and the second phase transition are close. 
Even if the energy scales are apart from each other as indicated by NANOGrav recent result~\cite{NANOGRAV:2023gor,NANOGrav:2023ctt,NANOGrav:2023hvm,NANOGrav:2023hfp,NANOGrav:2023icp}, the slope of the spectrum is different from the standard string spectrum in low frequency region~\cite{Dunsky:2021tih,Maji:2023fba,Lazarides:2023ksx}. The string-wall spectrum predicted by Nambu-Goto description fits the NANOGrav result well~\cite{Maji:2023fba,Lazarides:2023ksx}. In our model, string-walls behave as strings with the winding number $\pm 2$ at the late time, and therefore, there may be an additional component at low frequency region of gravitational wave spectrum in addition to the string-wall spectrum predicted by~\cite{Dunsky:2021tih}. We will leave the examination in the field-theoretical simulation based on our model as a future work.

We note a few comments from other observational point of view.
Though there exists a massless mode in our model as an outcome of the symmetry breaking, in the guaged theories by which we are motivated, the massless mode will be eaten by a gauge field and becomes massive as the longitudinal mode of the gauge field. In addition, the other energy scales we assume are high enough. Therefore, our (gauged) model does not contain dark radiation and  meets the current observation of cosmic microwave background, e.g., no deviation from the standard $N_{\rm eff}$ will be seen. 
To avoid the particle decay into Standard Model particles (photons), %
the interaction of these fields and Standard Model needs to be small enough. 
Quasi-Nambu-Goldston boson $m_{\rm qNG}\sim 10^2 {\rm GeV}$ and the second radial mode $m_2> m_{\rm qNG}$ can be dark matter candidates. Emission of such particles from string-walls will contribute enough to the dark matter abundance since the energy scales of the formation of the walls and collapse of the string-walls are not far from those in a single-axion model in \cite{Hiramatsu:2012gg}. For the concrete predictions require further analyses and numerical simulations.

\begin{acknowledgments}
We would like to thank Tsutomu Yanagida and Graham White for fruitful discussion. 
This work is supported by JSPS KAKENHI Grant Numbers JP22H01221 (ME), JP22K20365 (YS), JP21K03559 and JP23H00110 (TH), and MEXT KAKENHI Grant Numbers JP20H00225 and JP20H02066 (IS).
The computation was carried out using General Projects on supercomputer ``Flow'' at Information Technology Center, Nagoya University and Yukawa Institute Computer Facility.
\end{acknowledgments}

\appendix
\section{Mass spectrum}
\label{app:mass}
We find the mass spectrum of our model. The vacuum solution is $|\zeta|=v_1$ and Eq.~\eqref{eq:vac_cond} under the hierarchy \eqref{eq:hierarchy}. There are four states: two radial modes, $\delta |\zeta|$ and $\delta |\phi|$, and two angular modes, $v_1\delta \alpha$ and $v_2\delta \beta$. There are no mixing between the radial modes and the angular modes. By diagonalizing the mass matrix for the radial modes, we obtain
\begin{widetext}
\begin{align}
    m_{1,2}^2=
    2 mv_1+\lambda_1 v_1^2+\lambda_2 v_2^2
    \pm \frac{\sqrt{8 m^3 v_1+4 \lambda_2 m^2
    \left(v_1^2+v_2^2\right)+4 \lambda_2 m v_1
    \left(\lambda_2 v_2^2-\lambda_1
    v_1^2\right)+\lambda_2 \left(\lambda_1
    v_1^2-\lambda_2
    v_2^2\right)^2}}{\sqrt{\lambda_2}}  \ ,
\end{align}
\end{widetext}
where the plus sign for $m_1^2$ and the minus sign for $m_2^2$ in the right-hand side.
Under the hierarchy, $m_1^2 = 2\lambda_1 v_1^2$.
In the limit $\tilde{m}=m/m_\ast\ll 1$, $m_2^2= 2\lambda v_2^2$. In opposite limit, $\tilde{m}\gg 1$, $m_2^2=4m v_1$. Similarly, by the diagonalization of the mass matrix for the angular modes, we find a mode with
\begin{align}
    m_3^2=m\left(4 v_1+\frac{v_2^2}{v_1}+\frac{2m}{\lambda_2}\right)\ ,
\end{align}
and a massless mode. Under the hierarchy, $m_3=4mv_1$. The massless mode is nothing but the NG boson along the flat direction on the torus (the torus knot). The massive mode with $m_3$ is understood as the quasi-NG boson orthogonal to the torus knot on the torus in the weakly interacting regime.  We note $m_2=m_3$ in the weakly interacting regime.

\section{Field configurations}
\label{app:field_config}

\subsection{Domain-wall solution}
We find wall solutions in our model. We note that these walls are transient objects because the temporal $Z_2$ as a center of $U(1)$ is not stable.

To find a solution describing the wall, we consider $\alpha$-constant surface in $XYZ$-space, namely, $\tilde{\phi}$ plane, where the wall has the minimal energy.

For the weakly interacting regime,  the lowest energy path between the two vacua is along the half circle with a radius $v_2$ on $\tilde{\phi}$-plane.~\footnote{More precisely, it is along the ellipse described by Eq.~\eqref{eq:weak_ellipse}.} The energy density is reduced to
\begin{align}
    E(\tilde{\phi})= \frac{1}{2}v_2^2 (\partial_x \varphi)^2 + m v_1 v_2^2 \cos(2\varphi) \ , 
\end{align}
where we assume a planer wall located at $x=0$ in three-dimensional space and $\varphi = \arg (\tilde{\phi})$. The variation gives 
\begin{align}
    -\partial_x\partial_x \varphi - 2 mv_1 \sin(2\varphi)=0 \ , 
\end{align}
which becomes 
\begin{align}
    -\partial_{\bar{x}}\partial_{\bar{x}} \bar{\varphi} +  \sin(\bar{\varphi})=0 \ , 
\end{align}
by defining 
\begin{align}
    \bar{x}=2 \sqrt{m v_1} x \ ,
\end{align}
\begin{align}
    \bar{\varphi}= 2\varphi \pm \pi \ .
\end{align}
This equation is a well-known sine-Gordon equation solved by
\begin{align}
   \bar{\varphi} = 4 \arctan (e^{\bar{x}}) \ , 
\end{align}
and thus,
\begin{align}
    \varphi = 2 \arctan (e^{2\sqrt{m v_1}x}) \mp \frac{\pi}{2}\ , 
\end{align}
which has a range of $\var\phi \in [-\pi/2, \pi/2]$ and $\var\phi \in [\pi/2, 3\pi/2]$. The former corresponds to the cyan dashed curve and the latter corresponds to the pink dashed curve in Fig.~\ref{fig:pot_torus_2_weak}.
Therefore, the wall width is 
\begin{align}
    \delta^{\rm (weak)}_{\rm DW}    \sim (2\sqrt{|m| v_1})^{-1}   \ .
\end{align}
By integrating the energy density with this solution, we obtain the wall tension
\begin{align}
    \sigma= \int_{-\infty}^{\infty} dx \, (E(\phi)-E(\pm i v_2)) = 4\sqrt{m v_1} v_2^2 \ .
\end{align}

For the strongly interacting regime, the lowest energy path between the two vacua is along the imaginary axis on $\tilde{\phi}$-plane. The energy density is written as
\begin{align}
    E(\tilde{\phi})=\frac{1}{2}(\partial_x \tilde{\phi}_I)^2 
    + \frac{\lambda_2}{4}(\tilde{\phi}^2_I-v_2^2)^2-mv_1\tilde{\phi}_I^2 \ ,
\end{align}
where we assume a planer wall located at $x=0$ in three-dimensional space and $\tilde{\phi}_I = \Im \tilde{\phi}$. The variation with respect to $\tilde{\phi}_I$ gives
\begin{align}
    -\partial_x\partial_x \tilde{\phi}_I + \lambda_2 \tilde{\phi} _I^3 -\lambda_2 \tilde{v}_2^2 \tilde{\phi}_I = 0 \ ,
\end{align}
which is reduced to 
\begin{align}
    -\partial_{\bar{x}} \partial_{\bar{x}} \bar{\phi}_I + \bar{\phi}_I^3 -  \bar{\phi}_I = 0 \ ,
\end{align}
by defining 
\begin{align}
    \bar{x} = \sqrt{\lambda} \tilde{v}_2 x\ ,
\end{align}
\begin{align}
    \bar{\phi}_I = \frac{\tilde{\phi}_I}{\tilde{v}_2} \ .
\end{align}
This equation has the solution,
\begin{align}
    \bar{\phi}_I= \tanh \Bigl(\frac{\bar{x}}{\sqrt{2}}\Bigr)\ , 
\end{align}
and thus,
\begin{align}
    \tilde{\phi}_I =  \tilde{v}_2 \tanh\Bigl(\frac{\sqrt{\lambda_2}\tilde{v}_2}{\sqrt{2}}x\Bigr) \ .
\end{align}
Therefore, the wall width is 
\begin{align}
    \delta^{\rm (strong)}_{\rm DW}\sim(\sqrt{\lambda_2} \tilde{v}_2)^{-1}   \ .
    \label{eq:delta_dw_strong}
\end{align}
By integrating the energy density with the solution, we obtain the wall tension
\begin{align}
    \sigma= \int_{-\infty}^{\infty} dx \, (E(\phi)-E(\pm i \tilde{v}_2)) = \frac{2 \sqrt{2}}{3} \sqrt{\lambda_2} \tilde{v}_2^3 \ .
\end{align}

\subsection{$\phi$-string solution}

Let us focus on a $\phi$-string:
\begin{align}
    \zeta = v_1 \ , 
\end{align}
and
\begin{align}
    \phi = g(r) e^{i l \Theta} \ ,
\end{align}
where $l$ is an integer.
The field equation of the $\phi$-field is 
\begin{align}
    &\frac{d^2 g}{dr^2}+\frac{1}{r}\frac{dg}{dr}-\frac{l^2}{r^2}g -\lambda_2 g (g^2-v_2^2)\nonumber \\ 
    &- 2 m v_1 g \cos (2 l \Theta)= 0 \ .
    \label{eq:phi_weak}
\end{align}
In the weakly interacting regime, the last term in the right-hand side of \eqref{eq:phi_weak} can be neglected. Then, the approximate solution to \eqref{eq:phi_weak} becomes
\begin{align}
    g(r) = v_2 \left[1- \left( \frac{l}{\sqrt{2\lambda_2} v_2 r}\right)^2 \right] \ .
\end{align}
From the observation, we find the string core size is 
\begin{align}
    \delta^{\rm (weak)}_{\phi} \sim (\sqrt{\lambda_2} v_2)^{-1}\ .
    \label{eq:delta_phi_weak}
\end{align}
Since the walls in the strongly interacting regime can be thought as extended $\phi$-strings between $\zeta$-strings, it is natural that \eqref{eq:delta_phi_weak} and \eqref{eq:delta_dw_strong} coincide with each other in the saturated limit~\eqref{eq:saturate_m}.

\section{Initial conditions}
\label{App:Initial}

We explain the initial conditions used in our cosmological numerical simulation.
We rescale the field variables in Eqs.~\eqref{eq:zeta} and \eqref{eq:phi} as
\begin{align}
    \zeta \rightarrow \frac{\bar{\zeta}}{a} \ , \quad 
    \phi \rightarrow \frac{\bar{\phi}}{a} \ .
\end{align}
Then, we absorb constants as 
\begin{align}
    \zeta \rightarrow \tilde{\zeta} = \sqrt{\lambda_1}\bar{\zeta} \ , \quad 
    \phi \rightarrow \tilde{\phi} = \sqrt{\lambda_2}\bar{\phi} \ .
\end{align}
We normalize the fields and the spacetime by $\sqrt{\lambda_1} v_1$ and obtain
\begin{align}
     \ddot{Z} =\delta^{ij} D_iD_j Z 
     - \Bigl( Z (|Z|^2-a^2 ) +a M P_2  \Phi^2
      +\frac{1}{3}%
      \frac{a^2 T^2}{v_1^2}Z \Bigr) \ ,  
\end{align}
and
\begin{align}
     \ddot{\Phi} =& \delta^{ij} D_iD_j \Phi
     - \Bigl( \Phi(|\Phi|^2-a^2 \frac{P_1}{P_2}R) +2 a M^\ast P_1 Z \Phi^\ast \nonumber \\
     & \qquad \qquad \qquad +\frac{P_1}{3P_2}%
     \frac{a^2 T^2}{v_1^2}\Phi\Bigr)\ , 
\end{align}
where
\begin{align}
    \bar{\zeta} = v_1 Z \ ,  \quad \bar{\phi} = \sqrt{\frac{P_2}{P_1}}v_1 \Phi \ , 
\end{align}
\begin{align}
    P_1 = \frac{1}{\lambda_1} \ , \quad P_2 = \frac{1}{\lambda_2} \ , \quad 
    M = \frac{m}{ v_1} \ , \quad  R = \Bigl(\frac{v_2}{v_1}\Bigr)^2  \ .
\end{align}
We adopt finite temperature initial distribution for $Z_i=(\Re Z, \Im Z)$ and find
\begin{align}
    \langle Z_i(x) Z_j(y) \rangle=\delta_{ij} \int \frac{d^3k }{(2\pi)^3} \frac{n_k}{\omega_k}e^{i {\bf k}\cdot ({\bf x}-{\bf y})} \ ,
\label{eq:ZZ}
\end{align}
\begin{align}
    \langle \dot{Z}_i(x) \dot{Z}_j(y) \rangle=\delta_{ij} \int \frac{d^3k }{(2\pi)^3} n_k\omega_k e^{i {\bf k}\cdot ({\bf x}-{\bf y})} \ ,
\label{eq:dotZdotZ}
\end{align}
and
\begin{align}
    \langle \dot{Z}_i(x) Z_j(y) \rangle=0 \ ,
\label{eq:ZdotZ}
\end{align}
where
\begin{align}
    n_k= \frac{1}{e^{\omega_k/T}-1} \ , 
\end{align}
\begin{align}
    \omega_k = \sqrt{k^2 + m_{\rm eff}^2} \ ,
\end{align}
$m_{\rm eff}$ is the effective masses of $Z$,
\begin{align}
    m_{\rm eff}^2 = m^2_Z := -1 + \frac{1}{3}    T_{\rm ini}^2 \ ,
\end{align}
and  $T_{\rm ini}$ is the temperature when the initial conditions are set in $v_1$ unit.
In the momentum space, Eqs.~\eqref{eq:ZZ}--\eqref{eq:ZdotZ} are reduced to 
\begin{align}
    \langle Z_i (k) Z_j(k^\prime)\rangle = \frac{n_k}{\omega_k}(2\pi)^3 \delta^3({\bf k}+{\bf k^\prime}) \delta_{ij} \ ,
\end{align}
\begin{align}
    \langle \dot{Z}_i (k) \dot{Z_j}(k^\prime)\rangle =n_k \omega_k (2\pi)^3 \delta^3({\bf k}+{\bf k^\prime}) \delta_{ij} \ ,
\end{align}
and
\begin{align}
    \langle \dot{Z}_i(k) Z_j(k^\prime) \rangle=0 \ .
\end{align}
When we discretize them in a finite spacetime (replacing the delta functions with dimensionless delta function as $\delta({\bf k})\rightarrow \delta_k \times (L/(2\pi))^3 $), we obtain
\begin{align}
    \langle Z_i \rangle = 0 \ , \qquad   \langle \dot{Z}_i \rangle = 0 \ ,
\label{eq:average}
\end{align}
\begin{align}
    \langle |Z_{k}|^2 \rangle = \frac{n_k}{\omega_k}L^3 \ , 
\label{eq:deviation}
\end{align}
and
\begin{align}
     \langle |\dot{Z}_{k}|^2 \rangle = n_k \omega_k L^3 \ .
\label{eq:deviation_dot}
\end{align}
We generate the initial condition by the Gaussian distribution with the average,~\eqref{eq:average}, and the deviations,~\eqref{eq:deviation} and \eqref{eq:deviation_dot}.
We similarly generate the initial condition for $\Phi$ with 
\begin{align}
    m^2_\Phi = 
    \Bigl(-R + \frac{1}{3}
    T_{\rm ini}^2
    \Bigr)\frac{P_1}{P_2} \ .
\end{align}

\section{Reconnection of string walls}

We show the time-lapse snapshots interpolating Figs.~\ref{fig:TE_weak_e} and \ref{fig:TE_weak_f}. We focus on the string walls undergoing the reconnection.

\begin{figure}
    \centering
        \vspace*{-5mm}
    \centering
        \includegraphics[width=1 \linewidth, keepaspectratio]{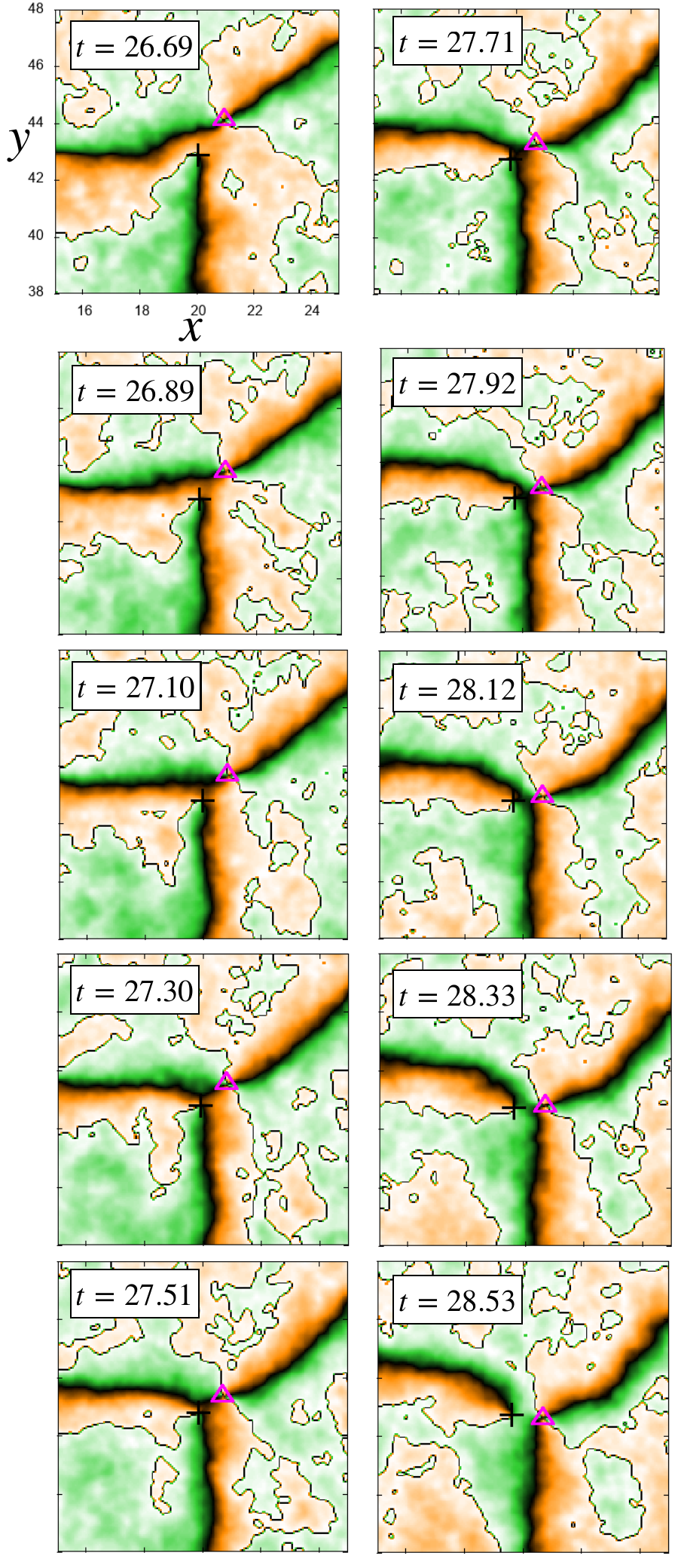}
    \caption{Enlarged figures of the string walls around $(x,y)=(20,40)$ in Fig.~\ref{fig:TE_weak} undergoing a reconnection. The schematic diagram is shown by Fig.~\ref{fig:weak_rec1}. The time range is from $t=26.69$ to $t=28.53$, between Figs.~\ref{fig:TE_weak_e} ($t=25.04$) and \ref{fig:TE_weak_f} ($t=29.17$).
    }
        \label{fig:s1_slow}
\end{figure}

\begin{figure*}
    \centering
        \vspace*{-5mm}
    \centering
        \includegraphics[width=0.5 \linewidth, keepaspectratio]{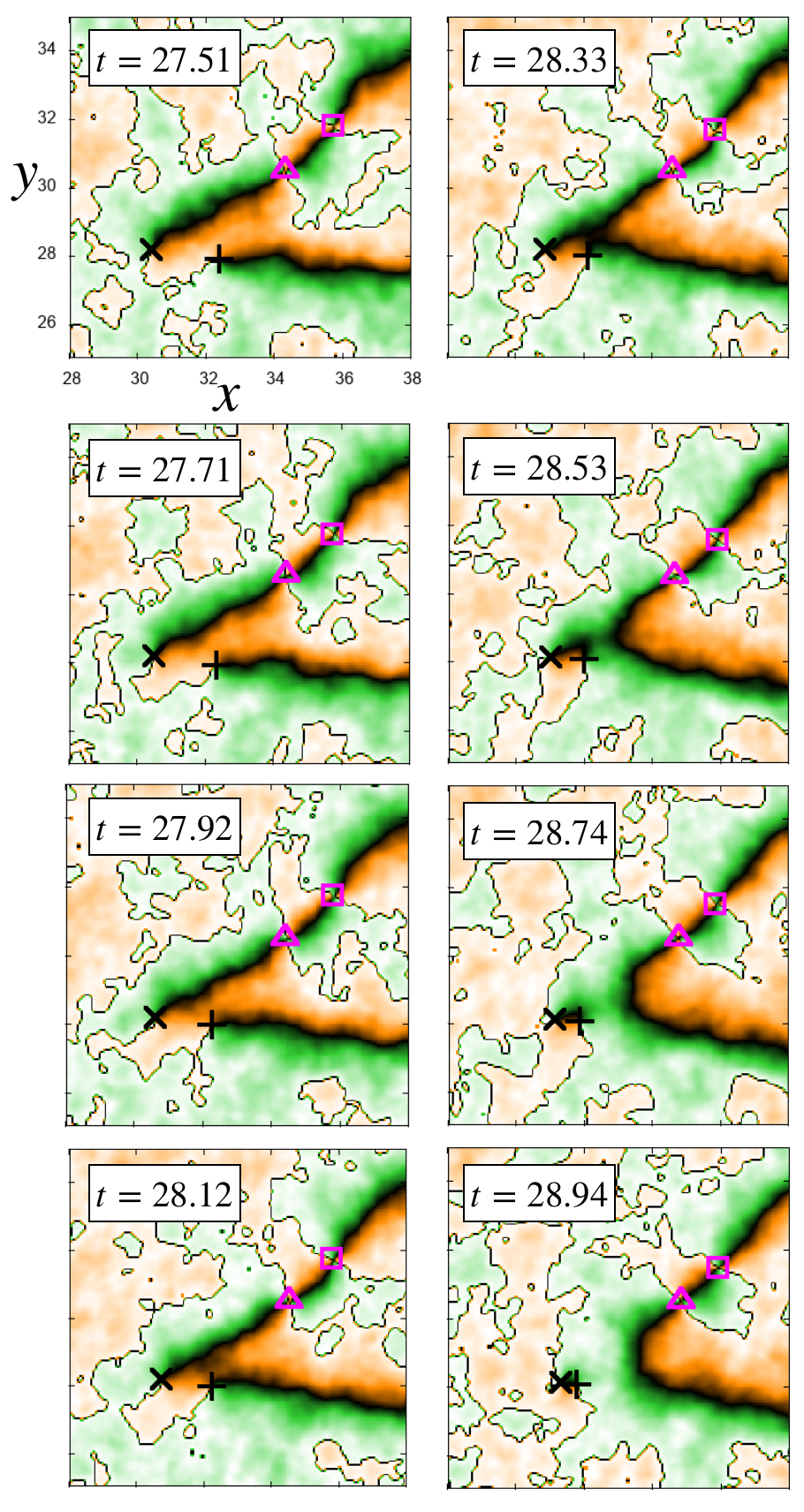}
    \caption{Enlarged figures of the string walls around $(x,y)=(35,30)$ in Fig.~\ref{fig:TE_weak} undergoing a reconnection. The schematic diagram is shown by Fig.~\ref{fig:weak_rec2}. The time range is from $t=27.51$ to $t=28.94$, between Figs.~\ref{fig:TE_weak_e} ($t=25.04$) and \ref{fig:TE_weak_f} ($t=29.17$).}
        \label{fig:s2_slow}
\end{figure*}

\clearpage

\bibliography{refs}
\end{document}